\documentclass[aps,superscriptaddress,amsmath,amssymb,floatfix,twocolumn,showpacs,amsfonts,longbibliography]{revtex4-1}
\usepackage{times}
\usepackage{threeparttable}
\usepackage{booktabs}
\usepackage{tabularx}
\usepackage{array}
\usepackage[varg]{txfonts}
\usepackage{textcomp}
\usepackage{graphicx}
\usepackage{subfigure}
\usepackage{mathtools}
\usepackage{color}
\usepackage[colorlinks=true,citecolor=blue,urlcolor=blue,linkcolor=blue,hyperindex]{hyperref}
\usepackage{braket}
\usepackage{overpic}
\usepackage{ulem}

\def\bk{{\bf k}}

\begin{document}

\title{Triply degenerate nodal lines in topological and non-topological metals}

\author{Zhihai Liu}
\affiliation{State Key Laboratory of Optoelectronic Materials and Technologies, School of Physics, Sun Yat-sen University, Guangzhou 510275, China}
\author{Luyang Wang}
\email{wangly@szu.edu.cn}
\affiliation{College of Physics and Optoelectronic Engineering, Shenzhen University, Shenzhen 518060, China}
\affiliation{State Key Laboratory of Optoelectronic Materials and Technologies, School of Physics, Sun Yat-sen University, Guangzhou 510275, China}
\author{Dao-Xin Yao}
\email{yaodaox@mail.sysu.edu.cn}
\affiliation{State Key Laboratory of Optoelectronic Materials and Technologies, School of Physics, Sun Yat-sen University, Guangzhou 510275, China}

\date{\today}

\begin{abstract}
  Topological nodal-line semimetals exhibit double or fourfold degenerate nodal lines, which are protected by symmetries. Here, we investigate the possibility of the existence of triply degenerate nodal lines in metals. We present two types of triply degenerate nodal lines, one topologically trivial and the other nontrivial. The first type is stacked by two-dimensional pseudospin-1 fermions, which can be viewed as an critical case of a tunable band-crossing line structure that contains a symmetry-protected quadratic band-crossing line and a non-degenerate band, and can split into four Weyl nodal lines under perturbations. We find that surface states of the nodal line structure are dependent on the geometry of the lattice and the surface termination. Such a metal has a nesting of Fermi surface in a range of filling, resulting in a density-wave state when interaction is included. The second type is a vortex ring of pseudospin-1 fermions. In this system, the pseudospins form Skyrmion textures, and the surface states are fully extended topological Fermi arcs so that the model exhibits 3D quantum anomalous Hall effect with a maximal Hall conductivity. The vortex ring can evolve into a pair of vortex lines that are not closed in the first Brillouin zone. A vortex line cannot singly exist in the lattice model if it is the only nodal feature of the system.
\end{abstract}

\maketitle

%%%%%%%%%%%%%%%%%%%%%%%%%%%%%%%%%%%%%%%%%%%%%%%%%%%%%%%%%%%%%%%%
\section{Introduction}\label{sec:intro}
%%%%%%%%%%%%%%%%%%%%%%%%%%%%%%%%%%%%%%%%%%%%%%%%%%%%%%%%%%%%%%%%
Topological semimetals (TSM) are systems which have symmetry-protected band crossings between the conduction and the valence bands in the Brillouin zone (BZ). The touching could be discrete points or lines which can yield zero-dimensional or one-dimensional Fermi surfaces, respectively. The well-know examples of touching points are Weyl and Dirac nodes which can be described by Weyl and Dirac Hamiltonian, respectively. In the Weyl TSM, a Weyl node can be regarded as a monopole in momentum space which carries a positive or negative chirality charge, with the net charge in the BZ being zero, so Weyl nodes must emerge in pairs with opposite chirality in crystals. The Dirac node is a fourfold degenerate point, which can be regarded as two overlapped Weyl nodes with opposite chirality.
%, in general, such two Weyl points appear at the same momentum being annihilate each other, therefore a Dirac semimetal system must host both time-reversal and inversion symmetry.
Weyl and Dirac TSMs exhibit exotic properties owing to the special band structures, such as Fermi-arc surface states~\cite{Xiangang.Wan2011,Su-Yang.Xu2015,Hongming.Weng2015PRX,
Shin-Ming.Huang2015,Mehdi.Kargarian2016} stretched between two Weyl points in the surface BZ and chiral anomaly~\cite{D.T.Son2013,Jun.Xiong2015,B.Z.Spivak2016,
Xiaochun.Huang2015,Cheng-Long.Zhang2016,Max.Hirschberger2016,Jennifer.Cano2017} in bulk. These systems have been intensively researched in both theoretical~\cite{A.A.Burkov2011PRL,Gabor.B.Halasz2012,S.M.Young2012,
Bohm-Jung.Yang2014,J.A.Steinberg2014,Chen.Fang2016,A.A.Burkov2016,N.P.Armitage2018} and experimental~\cite{Z.K.Liu2014NM,Z.K.Liu2014Science,Su-Yang.Xu2015,B.Q.Lv2015PRX,B.Q.Lv2015NP,L.X.Yang2015} communities. Beyond Weyl and Dirac nodes, TSM can also host three-~\cite{Weng.HongmingPRB2016,Zhu.ZimingPRX2016}, four-~\cite{Zhijun.Wang2012,Zhijun.Wang2013}, six-~\cite{Barry.Bradlyn2016} and eight-band~\cite{Wieder2016PRL} touching points which are protected by space group symmetries.

The systems with line-like band touching are commonly termed topological nodal line semimetals (TNLSM), and the symmetry-protected nodal lines exhibit various forms, such as a line running through the BZ but resetting at its boundary~\cite{Yuanping.Chen2015,Qi-Feng.Liang2016,C.Chen2017}, or a loop inside the BZ~\cite{Gang.Xu2011,Kieran.Mullen2015,Yige.Chen2016}, or even a chain~\cite{Tomas.Bzdusek2016,Rui.Yu2017} or a link~\cite{Wei.Chen2017,Zhongbo.Yan2017,Po-Yao.Chang2017}. Breaking the protecting symmetry, the nodal line evolves into a full gap or several nodal points. The nodal lines can exist in a system with or without spin-orbit coupling (SOC). When SOC is present, the system usually requires an additional symmetry, the glide symmetry for the nodal line to exist~\cite{ChenFang2015,Ding-Fu.Shao2018}. Based on the structure features of nodal lines, generally, we can continuously tune the parameters to deform the nodal line and even encounter a topological transition~\cite{Michael.Phillips2014,ChenFang2015,Lih-KingLim2017,Zhesen.yang2019}. A lot of theoretical efforts have been devoted~\cite{Petr.Horava2005,A.A.Burkov2011PRB,Youngkuk.Kim2015,Rui.Yu2015,
Yige.Chen2015,Y.H.Chan2016,Fang.Chen2016} and some realistic materials~\cite{Guang.Bian2016,Leslie.M.Schoop2016,C.Chen2017,C-J.Yi2018,Qinghui.Yan2018} have also been reported to realize the TNLSM phase. Nodal line systems also present some specific topological physics such as drumhead-like flat surface bands~\cite{Yige.Chen2015,Hongming.Weng2015PRB}.

A Dirac (Weyl) node in TSM can be regarded as a quasiparticle which corresponds to a massless relativistic Dirac (Weyl) fermion with spin-$1/2$ in the context of particle physics. In contrast, in condensed matter physics, the existence of fermions with higher pseudospin~\cite{Z.Lan2011,Barry.Bradlyn2016} is allowed due to lesser symmetry constraint. For example, the pseudospin-1 fermion can appear in the kagome lattice~\cite{Dmitry.Green2010} and triangular-kagome lattice~\cite{Wang2018}. In these systems, a flat energy band crosses the touching point and presents a triply degenerate pseudospin-$1$ fermion~\cite{J.L.Manes2012}.
When the Fermi level crosses the triply degenerate point, these systems present a 2D planar Fermi surface, so these systems are topological metals~\cite{Yan-Qing.Zhu2017}. Some other systems with topological triply degenerate points have also been investigated~\cite{Haiping.Hu2018PRL,Haiping.Hu2018PRA}. The pseudospin-$3/2$ fermion~\cite{Motohiko.Ezawa2016} is also allowed by space group symmetries, which is from a four-band touching. These four bands can be divided into two groups by their different Fermi velocities or helicities $3/2$ and $1/2$. Constructing a closed surface enclosing the touching point, we obtain the Chern numbers $\pm3$ and $\pm1$ for helicity $3/2$ and $1/2$ bands, respectively. Therefore, in contrast with the Weyl and pseudospin-$1$ fermion which has topological charge $\pm1$ and $\pm2$, respectively, the pseudospin-$3/2$ fermion has a total topological charge $\pm4$~\cite{Chang2017PRL,Peizhe.Tang2017}.

In this paper, we investigate the possibility of triply degenerate nodal lines in lattice systems, which may exhibit topological properties. We present two kinds of systems which possess such nodal features. Firstly, we consider a star lattice model whose band structures present a tunable band-crossing point (TBCP) at the BZ center. The TBCP includes a quadratic band-crossing point (QBCP) and a non-degenerate band, and can be tuned into a linearly dispersed three-band-crossing point (pseudospin-$1$ fermion) by fine tuning~\cite{Mengsu.Chen2012} (see Fig.~\ref{FigBS}). The QBCP is formed by a flat band touching the upper or lower dispersing band and is protected by time reversal symmetry and $C_6$ symmetry~\cite{Kai.Sun2009PRL,Tsai_2015}. Such a band structure also exists in some other 2D lattices, such as the Lieb lattice~\cite{R.Shen2010,Apaja2010PRA}, $\mathcal{T}_3$ lattice~\cite{Bercioux.D2009}, square-octagon lattice~\cite{Li.Jun2020}, stacked triangular lattice~\cite{Dora2011PRB} and the above-mentioned kagome and triangular-kagome lattices. In the $AA$-stacked 3D star lattice, when the interlayer hopping amplitudes on different sites are equal, the TBCP evolves into a tunable band-crossing line (TBCL), and we can obtain a three-band-crossing line by fine tuning, which we will call the pseudospin-1 nodal line (see Table~\ref{Nodes}). Although the pseudospin-1 nodal line is not protected by symmetries and is unstable under perturbations, we view it as a critical case which has a simple Hamiltonian and understanding which is helpful to understand a TBCL that is protected by symmetries. Adjusting the interlayer hopping amplitudes on different sites to break the $C_6$ symmetry down to $C_2$ symmetry, we find that the pseudospin-1 nodal line is split into four doubly degenerate Weyl nodal lines which are stacked by 2D Dirac nodes, and even if the parameters slightly deviate from those at the triple degeneracy, the four Weyl nodal lines still exist. For sufficiently small interlayer hoppings which can be considered as a perturbation, we derive a 3D effective Hamiltonian to describe the pseudospin-$1$ nodal line and the splitting. A remarkable feature of the pseudospin-1 nodal line system is its instability to density-wave state when interaction in included, which is insensible to filling. The other triply degenerate nodal line system investigated in this work is a pseudospin-1 vortex ring model, which exhibits 3D quantum anomalous Hall effect with a maximal Hall conductivity due to surface Fermi arcs wrapping around the full surface BZ. The vortex ring can evolve into a couple of vortex lines. However, such a vortex line cannot exist on its own in the lattice model, similar with the fermion doubling of Dirac (Weyl) fermions~\cite{H.B.Nielsen1981,OskarVafek2014}. We also study the Landau level structures and surface states of both systems. We find that the vortex ring model has topological surface states and hence is topological, while the AA-stacked star lattice has geometry-dependent surface states and is non-topological.
%%%%%%%%%%%%%%%%%%%%%%%%%%%%%%%%%%%%%%%%%%%
\renewcommand\arraystretch{1.2}
\begin{table}[b]
  \centering
  \caption{List of nodal-line and nodal-point terminologies used in this paper, and the degeneracy and dispersion of each one. The spin degree of freedom is not considered, and the dispersion for nodal lines refers to that in the direction perpendicular to the nodal line.}
  %\begin{threeparttable}
  \begin{tabular}{p{110 pt} >{\centering}p{70 pt}
  >{\centering\arraybackslash}p{50 pt}}
  \toprule[0.7pt] \toprule[0.7pt]
  \specialrule{0em}{0.8pt}{0.8pt}
  Nodal-line/point & Degeneracy & Dispersion \\
  \specialrule{0em}{0.8pt}{0.8pt}
  \midrule[0.5pt]
  \specialrule{0em}{0.8pt}{0.8pt}
  Pseudospin-1 nodal line\tnote{1} & Triple & Linear \\
  Pseudospin-1 vortex ring\tnote{2} & Triple & Linear \\
  Quadratic band-crossing line & Double & Quadratic \\
  Weyl nodal line & Double & Linear \\
  Pseudospin-1 fermion & Triple & Linear \\
  Quadratic band-crossing point & Double & Quadratic \\
  Dirac point & Double & Linear \\
  \specialrule{0em}{0.8pt}{0.8pt}
  \bottomrule[0.7pt] \bottomrule[0.7pt]
  \end{tabular}
   %\begin{tablenotes}
    %    \footnotesize
    %    \item[1] running through the whole BZ along $k_z$.
    %    \item[2] forming a ring in the BZ.
   %\end{tablenotes}
  %\end{threeparttable}
\end{table}\label{Nodes}
%%%%%%%%%%%%%%%%%%%%%%%%%%%%%%%%%%%%%%%%%%%

%%%%%%%%%%%%%%%%%%%%%%%%%%%%%%%%%%%%%%%%%%%%%%%%%%%%%%%%%%%%%%%%
\section{Pseudospin-1 nodal line}\label{sec:starL}
%%%%%%%%%%%%%%%%%%%%%%%%%%%%%%%%%%%%%%%%%%%%%%%%%%%%%%%%%%%%%%%%
\subsection{Pseudospin-1 fermions in 2D}\label{subsec:model}
\begin{figure}[t]
  \centering\includegraphics[width=3.4in]{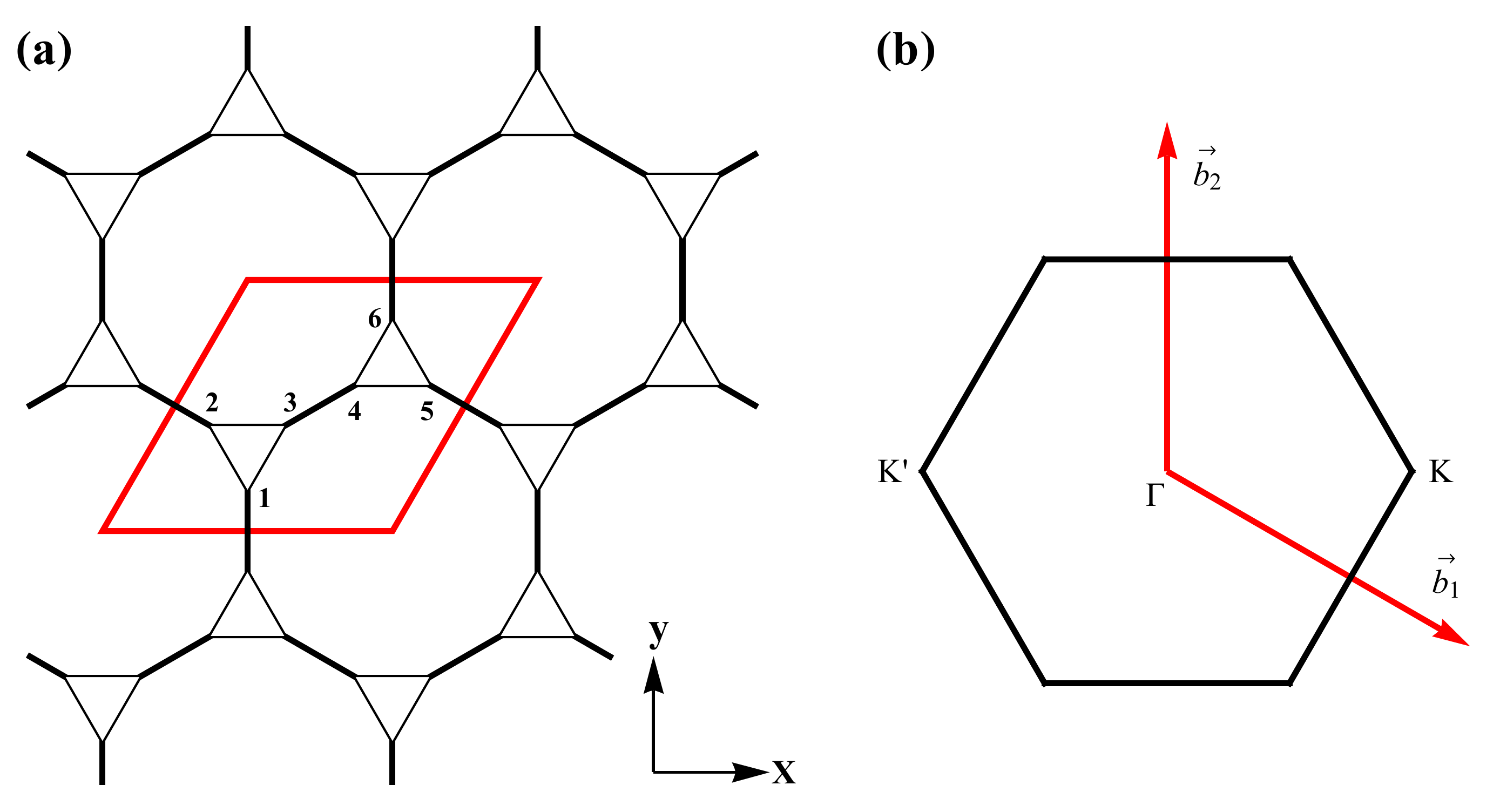}
  \caption{(a) The star lattice, the thick lines and thin lines represent the inter-triangle and intra-triangle hopping, respectively. The red parallelogram denotes a unit cell, with its six sites labeled. (b) The first Brillouin zone of the star lattice, with ${\bf b}_1$ and ${\bf b}_2$ representing the primitive reciprocal vectors.}\label{FigSL}
\end{figure}

The star lattice~\cite{Yan-Zhen.Zheng2007,Zewei.Chen2012} can be regarded as a honeycomb lattice with each site replaced by a triangle, as shown in Fig.~\ref{FigSL}(a). The red parallelogram denotes a unit cell with six sites that form six sublattices. We consider only one orbital on each site and assume the lattice constant $a=1$. The first BZ is shown in Fig.~\ref{FigSL}(b), in which ${\bf b}_1$ and ${\bf b}_2$ are the primitive vectors in the reciprocal space, and ${\bf \Gamma}$ labels the center of the BZ while ${\bf K}$ and ${\bf K^\prime}$ label two inequivalent corners.

%%%%%%%%%%%%%%%%%%%%%%%%%%%%%%%%%%%%%%%%%%%
%%%%%%%%%%%%%%%%%%%%%%%%%%%%%%%%%%%%%%%%%%%

\begin{figure}[t]
  \centering\includegraphics[width=3.4in]{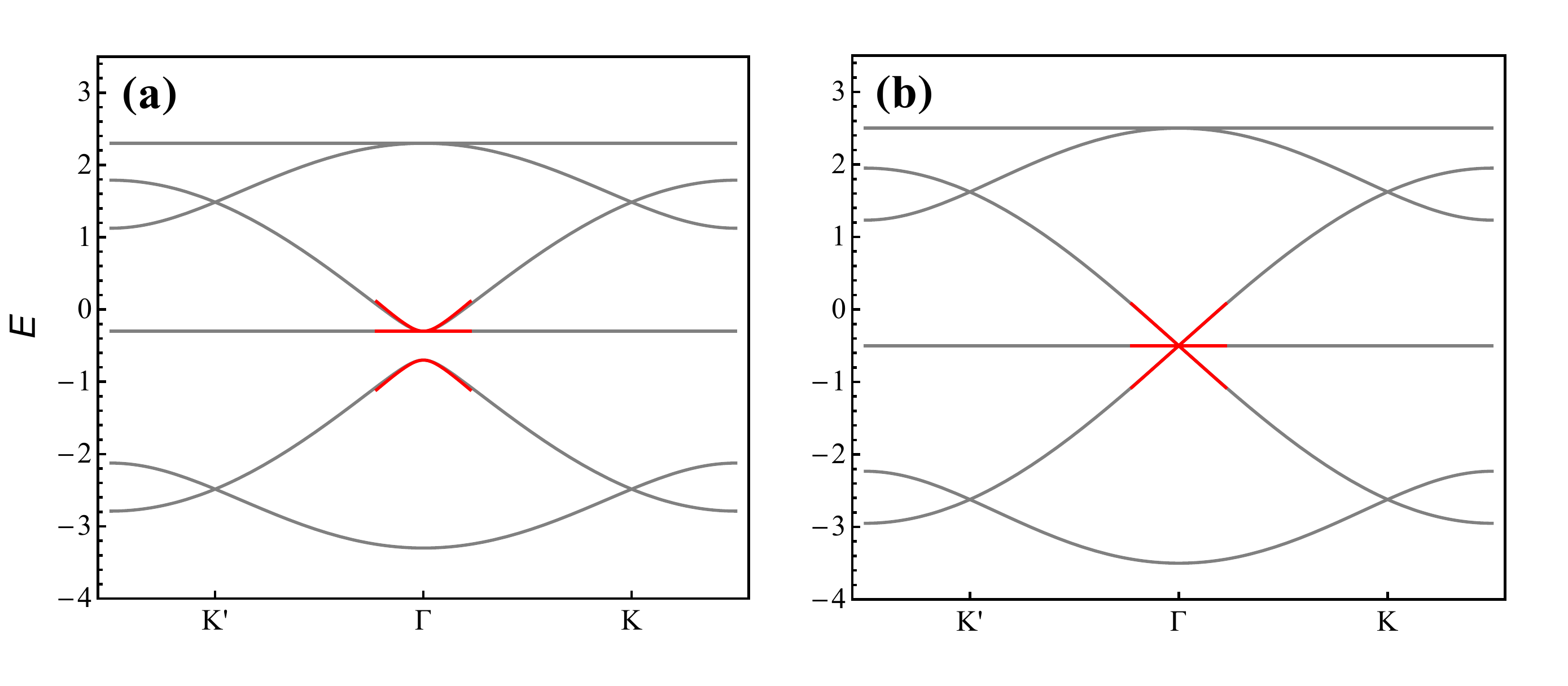}
  \caption{Band structures of the 2D star lattice with (a) $t^\prime=1.3t$ and (b) $t^\prime=t_c=1.5t$. The red lines depict the spectrum of the effective Hamiltonian.}\label{FigBS}
\end{figure}

We consider both the intra-triangle hopping and the nearest-neighbor inter-triangle hopping, with amplitude $t$ and $t^\prime$, respectively. The tight-binding Hamiltonian is
\begin{eqnarray}
\mathcal{H}_0 = - t \sum_{\mathclap{<ij>\sigma}} c^{\dagger}_{i\sigma} c_{j\sigma} - t^\prime \sum_{\mathclap{\ll ij\gg\sigma}} c^{\dagger}_{i\sigma} c_{j\sigma} + \mathrm{h.c.},
\label{Eq.LSH}
\end{eqnarray}
%%%%%%%%%%%%%%%%%%%%%%%%%%%%%%%%%%%%%%%%%%%
where $\mathrm{c}^{\dagger}_{i\sigma}$ $(\mathrm{c}_{i\sigma})$ creates (annihilates) an electron with spin $\sigma$ at site $i$, and $<ij>$ and $\ll ij\gg$ represent the intra-triangle hopping and the nearest-neighbor inter-triangle hopping, respectively. Taking the Fourier transformation, we get the Hamiltonian in momentum space $\mathcal{H}_0=\sum_{{\bf k}\sigma}{\Psi}_{\sigma}^{\dagger}({\bf k})H_0({\bf k}){\Psi}_{\sigma}({\bf k})$, where ${\Psi}_{\sigma}({\bf k})=(c_{1{\bf k}\sigma}, c_{2{\bf k}\sigma}, ..., c_{6{\bf k}\sigma})^T$, with $c_{i{\bf k}\sigma}$ annihilating an electron on sublattice $i$ with momentum $\bk$ and spin $\sigma$. We will not consider the spin degree of freedom until further specified. The band structure is obtained by diagonalizing $H_0({\bf k})$, as shown in Fig.~\ref{FigBS}, which contains two flat bands. By tuning the ratio $t^\prime/t$, the lower flat band touches a dispersing band either above or below it, realizing a TBCP. At a critical value $t^\prime=t_c\equiv1.5t$, a linear crossing point with three bands touching is achieved, forming a pseudospin-$1$ fermion. At high energies, Dirac points emerge at ${\bf K}$ and ${\bf K^\prime}$.

An effective Hamiltonian near the triple degeneracy can be derived by the $k\cdot p$ method~\cite{Wang2018}, which is
%%%%%%%%%%%%%%%%%%%%%%%%%%%%%%%%%%%%%%%%%%%
\begin{eqnarray}
H^{2D}_{eff}({\bf k}) = \epsilon_0+vk_x S_x+vk_y S_y,
\label{Eq.2DefH}
\end{eqnarray}
%%%%%%%%%%%%%%%%%%%%%%%%%%%%%%%%%%%%%%%%%%%
where $S_i$'s are three out of eight Gell-Mann matrices
%%%%%%%%%%%%%%%%%%%%%%%%%%%%%%%%%%%%%%%%%%%
\begin{eqnarray}
S_x=
\begin{pmatrix}
0 & 0 & i \\
0 & 0 & 0 \\
-i & 0 & 0
\end{pmatrix},
S_y=
\begin{pmatrix}
0 & 0 & 0 \\
0 & 0 & i \\
0 & -i & 0
\end{pmatrix},
S_z=
\begin{pmatrix}
0 & -i & 0 \\
i & 0 & 0 \\
0 & 0 & 0
\end{pmatrix},
\end{eqnarray}
%%%%%%%%%%%%%%%%%%%%%%%%%%%%%%%%%%%%%%%%%%%
which satisfy the angular momentum algebra $[S_i, S_j]=i\epsilon_{ijk}S_k$. The parameters are $\epsilon_0=-0.5t$ and $v=\sqrt{\tfrac{3}{8}}t$. When $t^\prime$ is away from $t_c$, the band structure can be described by adding an perturbation term
%%%%%%%%%%%%%%%%%%%%%%%%%%%%%%%%%%%%%%%%%%%
\begin{eqnarray}
\begin{pmatrix}
-\Delta & 0 & 0 \\
0 & -\Delta & 0 \\
0 & 0 & \Delta
\end{pmatrix},
\label{eq.delta}
\end{eqnarray}
%%%%%%%%%%%%%%%%%%%%%%%%%%%%%%%%%%%%%%%%%%%
where $\Delta=t^\prime-t_c$. Fig.~\ref{FigBS} shows the band structures for $t^\prime\ne t_c$ and $t^\prime=t_c$, where the spectrum of the 2D pseudospin-1 Hamiltonian is depicted in red lines.
%%%%%%%%%%%%%%%%%%%%%%%%%%%%%%%%%%%%%%%%%%
%%%%%%%%%%%%%%%%%%%%%%%%%%%%%%%%%%%%%%%%%%%

\subsection{Pseudospin-1 nodal line in 3D}\label{subsec:3DH}
\begin{figure}[t]
  \centering
  \subfigure{\includegraphics[width=1.6in]{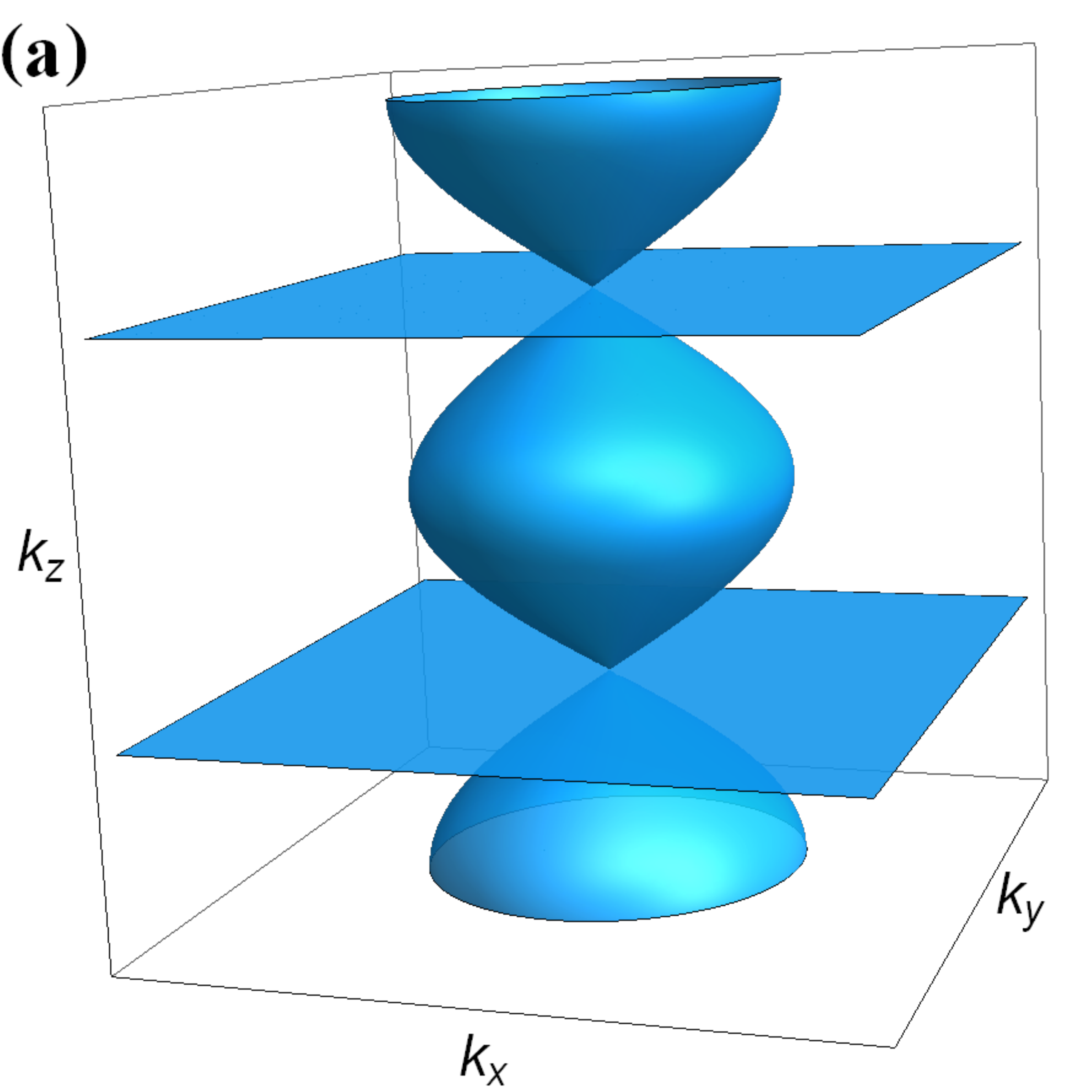}\label{SNa}}~~
  \subfigure{\includegraphics[width=1.6in]{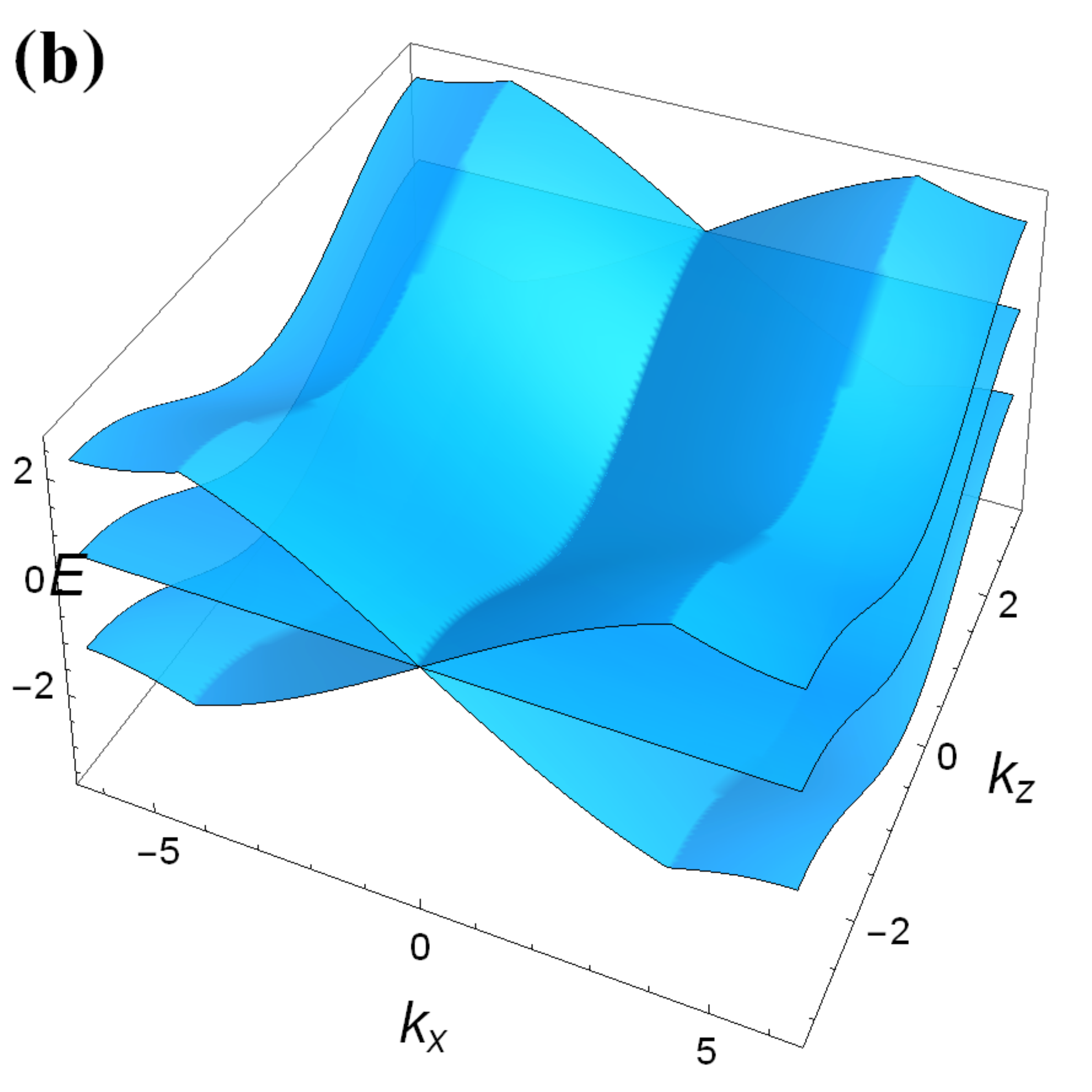}\label{SNb}}
  \caption{(a) The Fermi surface of the 3D star lattice at half filling. (b) The pseudospin-$1$  nodal line, with $k_y=0$.}
  \label{FigSN}
\end{figure}

Next, we consider a 3D lattice formed by $AA$-stacking of the 2D star lattice, which we call the 3D star lattice. The Hamiltonian in momentum space is obtained by introducing a $k_z$-dependent term
\begin{eqnarray}
\mathcal{H}_z=-\sum_{m{\bf k}\sigma}2t_z^m\cos k_z c^{\dagger}_{m{\bf k}\sigma} c_{m{\bf k}\sigma},
\label{Eq:Hz}
\end{eqnarray}
where $t^m_z$ represents the amplitude of the nearest-neighbor interlayer hopping at sublattice $m$ ($m=1, ... , 6$, see Fig.~\ref{FigSL}(a)). Now, we assume all sublattices have equal interlayer hopping, i.e. $t^m_z=t_z$, then we obtain the 3D band structure, which, at a fixed $k_z$ and $t^\prime=t_c$, has a pseudospin-$1$ fermion at $(0, 0, k_z)$ and Dirac points at the corners of the 2D slice of the 3D hexagonal BZ. Along the $k_z$-axis, the pseudospin-1 fermions form a pseudospin-1 nodal line. For $t^\prime$ deviating from $t_c$, the pseudospin-1 nodal line is replaced by a quadratic band-crossing nodal line and a non-degenerate band. Obviously, by adding a $k_z$-dependent term to the 2D Hamiltonian (\ref{Eq.2DefH}), we can obtain the 3D effective Hamiltonian for the pseudospin-1 nodal line,
%%%%%%%%%%%%%%%%%%%%%%%%%%%%%%%%%%%%%%%%%%%
\begin{equation}
H^{3D}_{eff}({\bf k}) = \epsilon_0+vk_x S_x+vk_y S_y - 2t_z \cos k_z.
\label{Eq:H3Deff}
\end{equation}
%%%%%%%%%%%%%%%%%%%%%%%%%%%%%%%%%%%%%%%%%%%
We show the Fermi surface (FS) at half filling in Fig.~\ref{FigSN}(a) and the pseudospin-1 nodal line in Fig.~\ref{FigSN}(b).

%%%%%%%%%%%%%%%%%%%%%%%%%%%%%%%%%%%%%%%%%%%
%%%%%%%%%%%%%%%%%%%%%%%%%%%%%%%%%%%%%%%%%%%

For a two-band-crossing node, we can define a winding number~\cite{Ahn2019PRX,Fuchs2010EPJB} $W_C=\frac{1}{2\pi}\alpha\oint_{C}d{\bf k}\cdot\nabla_{\bf k}\theta_{\bf k} = \alpha N_w$ where $\alpha=\pm1$ is a band index, $C$ an equienergy contour around the node and $\theta_{\bf k}$ the relative phase between two components of spinor wave-functions of the two-band Hamiltonian. We have also defined a positive (negative) $N_w$ as the number of counterclockwise (clockwise) rotations that a pseudospin vector undergoes when the eigenvector rotates one time around the node counterclockwise in the ${\bf k}$-parameter space~\cite{Park2011PRB}. In 2D, for the QBCP, the winding numbers are 2 and $-2$ with respect to the upper and lower band, respectively, while for the pseudospin-1 fermion, the winding numbers extracted from the pseudospin textures of the system (see Fig.~\ref{FigPSP}) are 1, 0 and $-1$ with respect to the upper, middle and lower band, respectively. Note that Dirac fermions also have winding number $\pm1$, but the implications of the winding number 1 for Dirac and pseudospin-1 fermions are different: the corresponding Berry phase is $\pi$ for Dirac fermions and 0 (modulo $2\pi$) for pseudospin-1 fermions. In 3D, evidently, on a contour which encloses a nodal line we can get the same winding numbers as the corresponding 2D nodal-point system, since the 3D system can be regarded as a stacking of the 2D systems along $k_z$.

\begin{figure}[t]
  \centering
  \subfigure{\includegraphics[width=1.6in]{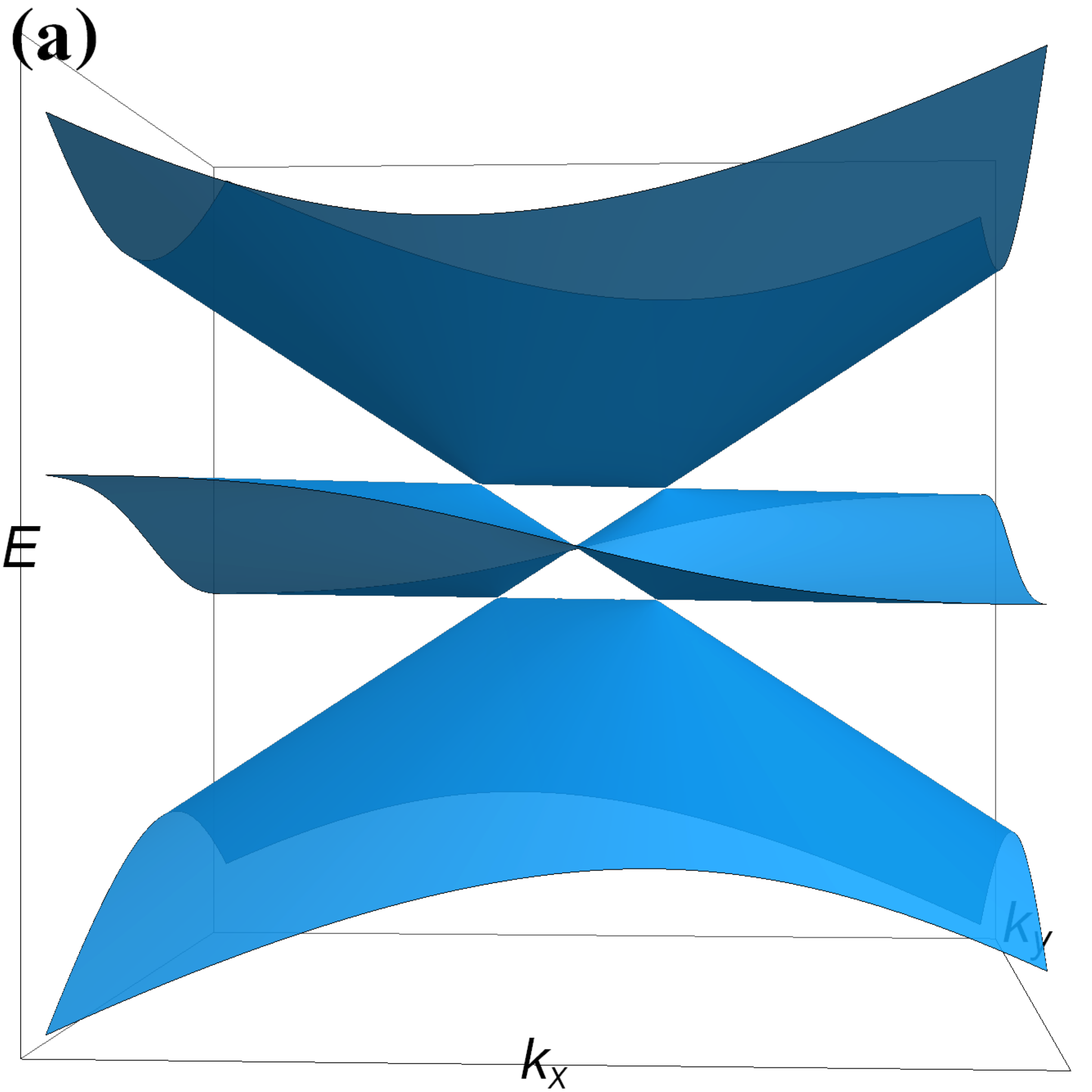}\label{BSa}}~~
  \subfigure{\includegraphics[width=1.6in]{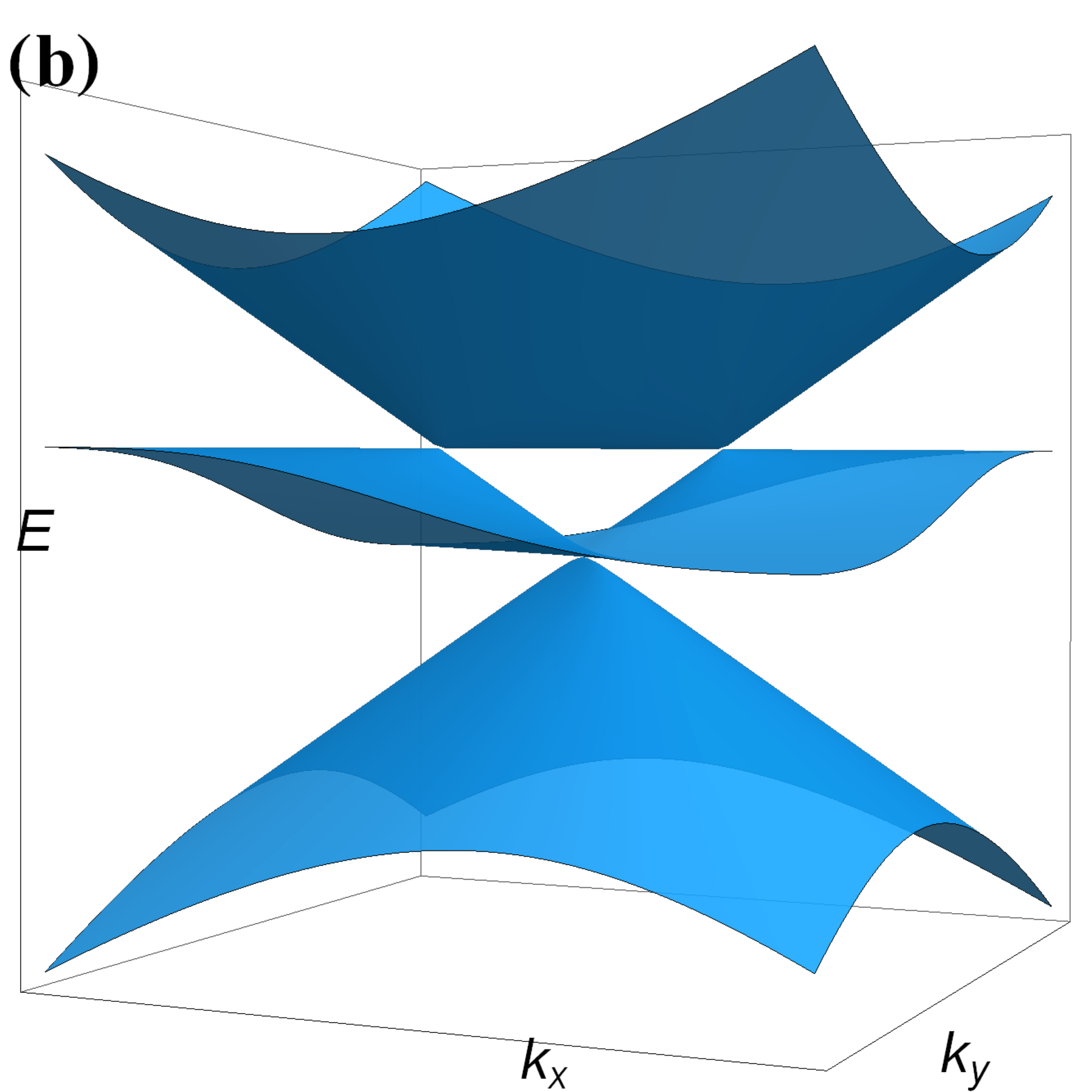}\label{BSb}}
  \subfigure{\includegraphics[width=1.6in]{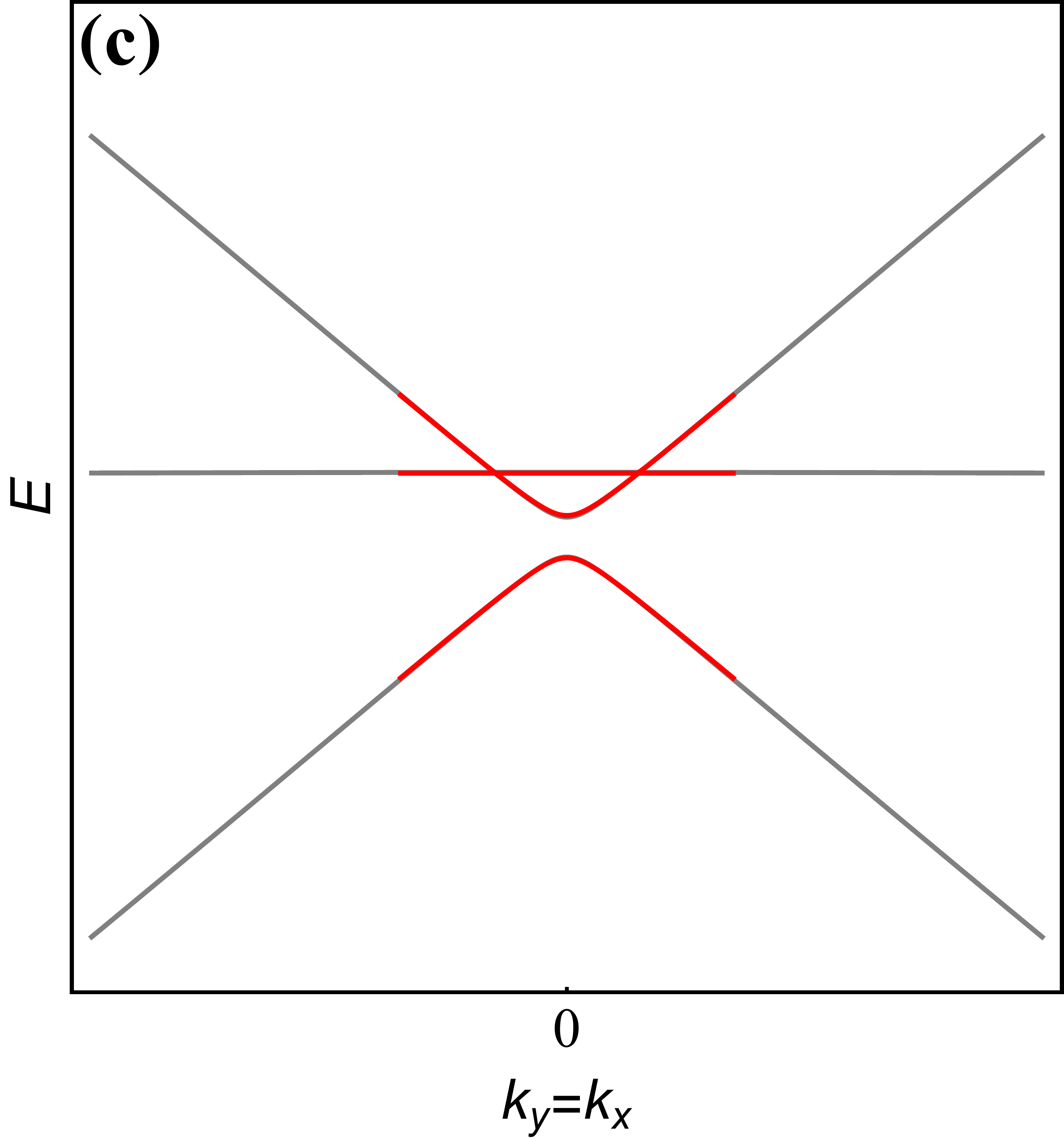}\label{BSc}}~~
  \subfigure{\includegraphics[width=1.6in]{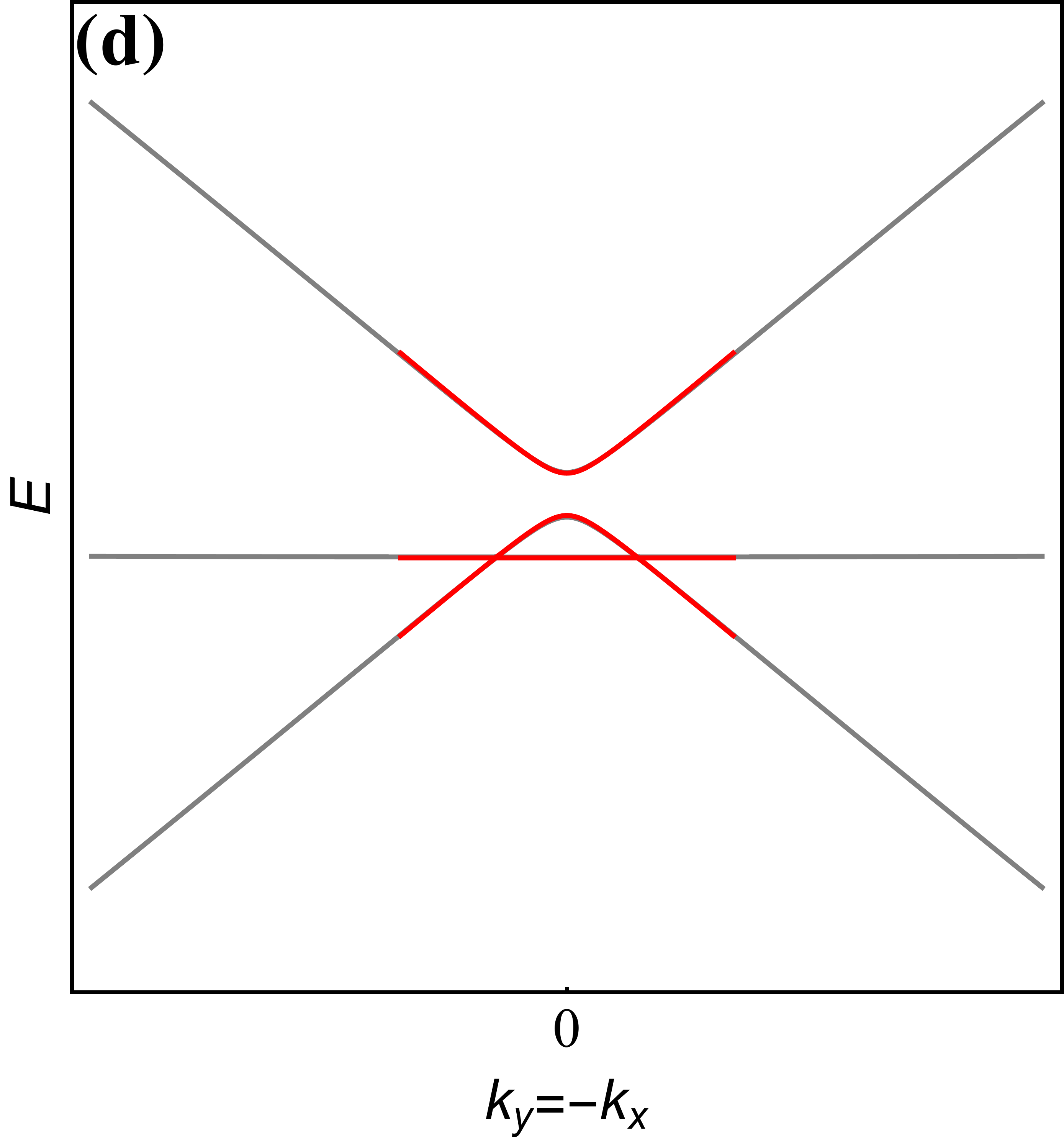}\label{BSd}}
  \caption{Schematic pictures of the splitting of the pseudospin-$1$ nodal line in the presence of $H'(k_z)$. (a, b) Dispersion in the $k_z=0$ plane with $\Delta=0$ and $\Delta=\Delta_a$, respectively. (c, d) Dispersion for $\Delta=0$ in the $k_z=0$ plane along $k_y=k_x$ and $k_y=-k_x$, respectively. The red lines represent the spectrum of the effective Hamiltonian. The inter-layer hopping parameters are $t^1_z=0.05t$, $t^2_z=0.25t$ and $t^3_z=0.15t$.} \label{FigSP}
\end{figure}

The pseudospin-1 nodal line may split into Weyl nodal lines under perturbations. For instance, in our 3D star lattice, when not all $t^i_z$'s are equal, but inversion symmetry is still conserved, i.e. $t^6_z=t^1_z$, $t^5_z=t^2_z$ and $t^4_z=t^3_z$ (see Fig.~\ref{FigSL}(a)), the pseudospin-$1$ nodal line splits into four Weyl nodal lines, as shown in Fig.~\ref{FigSP}(a) for $k_z=0$. If a perturbation $\Delta\neq0$ is considered via Eq.~(\ref{eq.delta}), when $\Delta_-<\Delta<\Delta_+$ where $\Delta_{\pm}=\tfrac{1}{6} \left(9\pm\sqrt{u_1^2+u_2^2+u_3^2-u_1u_2-u_1u_3-u_2u_3}\right)$, the four Weyl nodal lines still exist. Fig.~\ref{FigSP}(b) shows the case with $\Delta=\Delta_-$, in which there are two Weyl nodal lines and a band-crossing line with quadratic dispersion along $k_y=-k_x$ and linear dispersion along $k_y=k_x$. The Weyl nodal lines are formed in two orthogonal directions in momentum space. For clarity, in Figs.~\ref{FigSP}(c) and~\ref{FigSP}(d) we plot the dispersion along $k_y=k_x$ and $k_y=-k_x$ in $k_z=0$ plane, respectively. Such a splitting can also occur in a Lieb-kagome lattice~\cite{Lim2020PRB}. If the interlayer hopping amplitudes are small and can be treated as a perturbation, we get a 3D effective Hamiltonian with the following term in place of the last term of Eq.~(\ref{Eq:H3Deff}),
%%%%%%%%%%%%%%%%%%%%%%%%%%%%%%%%%%%%%%%%%%%
\begin{eqnarray}
H^{\prime}(k_z)=
\begin{pmatrix}
-\tfrac{u_1}{2}-\tfrac{u_2}{2} & \tfrac{u_1}{2\sqrt{3}}-\tfrac{u_2}{2\sqrt{3}} & 0 \\
\tfrac{u_1}{2\sqrt{3}}-\tfrac{u_2}{2\sqrt{3}} & -\tfrac{u_1}{6}-\tfrac{u_2}{6}-\tfrac{2u_3}{3} & 0 \\
0 & 0 & -\tfrac{u_1}{3}-\tfrac{u_2}{3}-\tfrac{u_3}{3}
\end{pmatrix}
\end{eqnarray}
%%%%%%%%%%%%%%%%%%%%%%%%%%%%%%%%%%%%%%%%%%%
where $u_i=2t_z^i\cos k_z$ for $i=1,2,3$. The spectra of the effective Hamiltonian with this perturbation are shown in Fig.~\ref{FigSP} by the red lines.

%%%%%%%%%%%%%%%%%%%%%%%%%%%%%%%%%%%%%%%%%%%
%%%%%%%%%%%%%%%%%%%%%%%%%%%%%%%%%%%%%%%%%%%

The splitting into four Weyl nodal lines can be understood in the following way. As mentioned above, the pseudospin-1 nodal line has winding number 1, 0 and $-1$ with respect to the upper, middle and lower band, respectively. We argue that the winding number $\pm1$ of pseudospin-1 fermions is equivalent to winding number $\pm2$ of QBCPs rather than $\pm1$ of Dirac fermions, for which the reason is that the Berry phase (modulo $2\pi$) for pseudospin-1 fermions and QBCPs are both $2\pi$, but for Dirac fermions it is $\pi$ ~\cite{Edward2006PRL,Yan-Qing.Zhu2017}. After the splitting, for the two Weyl nodal lines between the upper and middle band, the calculation gives winding number 1 with respect to the upper band and $-1$ with respect to the middle band, and for the two Weyl nodal lines between the lower band and middle band, the calculation gives winding number $-1$ with respect to the lower band and 1 with respect to the middle band. Therefore, we can say that the winding number of each band is conserved. This is similar to the splitting of QBCP~\cite{Gail2012PRB}.

\subsection{Landau level structure}\label{subsec:LLs}
Here we study the Landau levels (LLs) of the TBCL in a magnetic field. The canonical momentum should be replaced by the gauge-invariant kinetic momentum, i.e. ${\bf P}\rightarrow {\bf \Pi}={\bf P}+e{\bf A}({\bf r})$ where ${\bf P}=\hbar {\bf k}$. We use the vector potential ${\bf A}({\bf r})=(-By,0,0)$ which generates a homogeneous magnetic field ${\bf B}$ along $z$-direction. Since ${\bf r}$ and ${\bf P}$ satisfy the commutation relations $[r_i,P_j]=\delta_{ij}i\hbar$ and $[r_i,r_j]=[P_i,P_j]=0$ where $i,j=x,y$, defining $l_B=\sqrt{\tfrac{\hbar}{eB}}$ as the magnetic length, we obtain $[\Pi_x,\Pi_y]=-i\tfrac{\hbar^2}{l^2_B}$. Then we introduce the ladder operators $\hat{a}=\tfrac{l_B}{\sqrt{2}\hbar}(\Pi_x-i\Pi_y)$ and $\hat{a}^{\dagger} = \tfrac{l_B} {\sqrt{2}\hbar} (\Pi_x+i\Pi_y)$, which satisfy the commutation relation $[\hat{a},\hat{a}^{\dagger}]=1$. In terms of the ladder operators, the Hamiltonian in a magnetic field can be written as
%%%%%%%%%%%%%%%%%%%%%%%%%%%%%%%%%%%%%%%%%%%
\begin{eqnarray}
H^{{\bf B}}_{eff}=\epsilon_{k_z} + \tfrac{v} {\sqrt{2}l_B}
\begin{pmatrix}
0 & 0 & i(\hat{a}^{\dagger}+\hat{a}) \\
0 & 0 & (\hat{a}^{\dagger}-\hat{a}) \\
-i(\hat{a}^{\dagger}+\hat{a}) & -(\hat{a}^{\dagger}-\hat{a}) & \delta
\end{pmatrix},
\label{Eq.MHS}
\end{eqnarray}
%%%%%%%%%%%%%%%%%%%%%%%%%%%%%%%%%%%%%%%%%%%
where $\epsilon_{k_z}=\epsilon_0-2t_z\cos k_z-\Delta$ and $\delta=2\sqrt{2}l_B\Delta/v$, and we have set $\hbar=1$. Let us temporarily neglect the $\epsilon_{k_z}$ term and assume the eigenvalues and eigenstates of Hamiltonian (\ref{Eq.MHS}) are $E_n$ and ${\bf \psi}_n$, respectively, where ${\bf \psi}_n=(u_{1n}, u_{2n}, u_{3n})^T$ is a three-spinor wave-function. They satisfy the eigenequation $H^{{\bf B}}_{eff}{\bf \psi}_n=E_n{\bf \psi}_n$. Explicitly, we have
%%%%%%%%%%%%%%%%%%%%%%%%%%%%%%%%%%%%%%%%%%%
\begin{equation}
\begin{split}
&\tfrac{iv}{\sqrt{2}l_B} (\hat{a}^{\dagger}+\hat{a})u_{3n}=E_nu_{1n},  \\
&\tfrac{v}{\sqrt{2}l_B} (\hat{a}^{\dagger}-\hat{a})u_{3n}=E_nu_{2n},   \\
&\tfrac{v}{\sqrt{2}l_B} \left[-i(\hat{a}^{\dagger}+\hat{a})u_{1n} - (\hat{a}^{\dagger}-\hat{a})u_{2n}\right] = (E_n-\delta)u_{3n}.
\end{split}
\label{eq.eqs}
\end{equation}
Substituting the first two equations to the third yields
%%%%%%%%%%%%%%%%%%%%%%%%%%%%%%%%%%%%%%%%%%%
\begin{eqnarray}
 \tfrac{v^2}{l^2_B}(2\hat{a}^{\dagger}\hat{a}+1)u_{3n}=E_n(E_n-\delta)u_{3n},
\end{eqnarray}
%%%%%%%%%%%%%%%%%%%%%%%%%%%%%%%%%%%%%%%%%%%
which has the solution $u_{3n}=c_n|n\rangle$ for $n\ge0$, where $|n\rangle$ is the eigenstate of the number operator $\hat{a}^\dagger \hat{a}$ with eigenvalue $n$ and $c_n$ is a coefficient depending on $n$, and
%%%%%%%%%%%%%%%%%%%%%%%%%%%%%%%%%%%%%%%%%%%
\begin{eqnarray}
E_n = \frac{v}{2l_B} \left(\delta + \lambda\sqrt{8n+4+{\delta}^2}\right),
\end{eqnarray}
%%%%%%%%%%%%%%%%%%%%%%%%%%%%%%%%%%%%%%%%%%%
where $\lambda=\pm 1$. Utilizing the first two of Eq.~(\ref{eq.eqs}) we can get the spinor wave-function
%%%%%%%%%%%%%%%%%%%%%%%%%%%%%%%%%%%%%%%%%%%
\begin{eqnarray}
{\bf \psi}_{n\lambda} =\frac{C_n}{\sqrt{2}\epsilon_n}
\begin{pmatrix}
i\lambda \left(\sqrt{n+1} |n+1\rangle + \sqrt{n} |n-1\rangle\right) \\
\lambda \left(\sqrt{n+1} |n+1\rangle - \sqrt{n} |n-1\rangle\right) \\
 \sqrt{2} \epsilon_n |n\rangle
\end{pmatrix},
\end{eqnarray}
%%%%%%%%%%%%%%%%%%%%%%%%%%%%%%%%%%%%%%%%%%%
where $C_n=\sqrt{\tfrac{1}{2}-\tfrac{\delta} {\sqrt{16n+8+\delta^2}}}$ and $\epsilon_n=\tfrac{l_B}{v}E_n$. We note that $E_n=0$ also is a solution of the eigenequation, with the corresponding eigenfunction
%%%%%%%%%%%%%%%%%%%%%%%%%%%%%%%%%%%%%%%%%%%
\begin{eqnarray}
{\bf \psi}_{n0}=\frac{1}{\sqrt{4n+6}}
\begin{pmatrix}
i\sqrt{n+1}|n+2\rangle - i\sqrt{n+2}|n\rangle  \\
 \sqrt{n+1}|n+2\rangle + \sqrt{n+2}|n\rangle   \\
0
\end{pmatrix}.
\end{eqnarray}
%%%%%%%%%%%%%%%%%%%%%%%%%%%%%%%%%%%%%%%%%%%
%%%%%%%%%%%%%%%%%%%%%%%%%%%%%%%%%%%%%%%%%%%
\begin{figure}[t]
  \centering
  \subfigure{\includegraphics[width=1.7in]{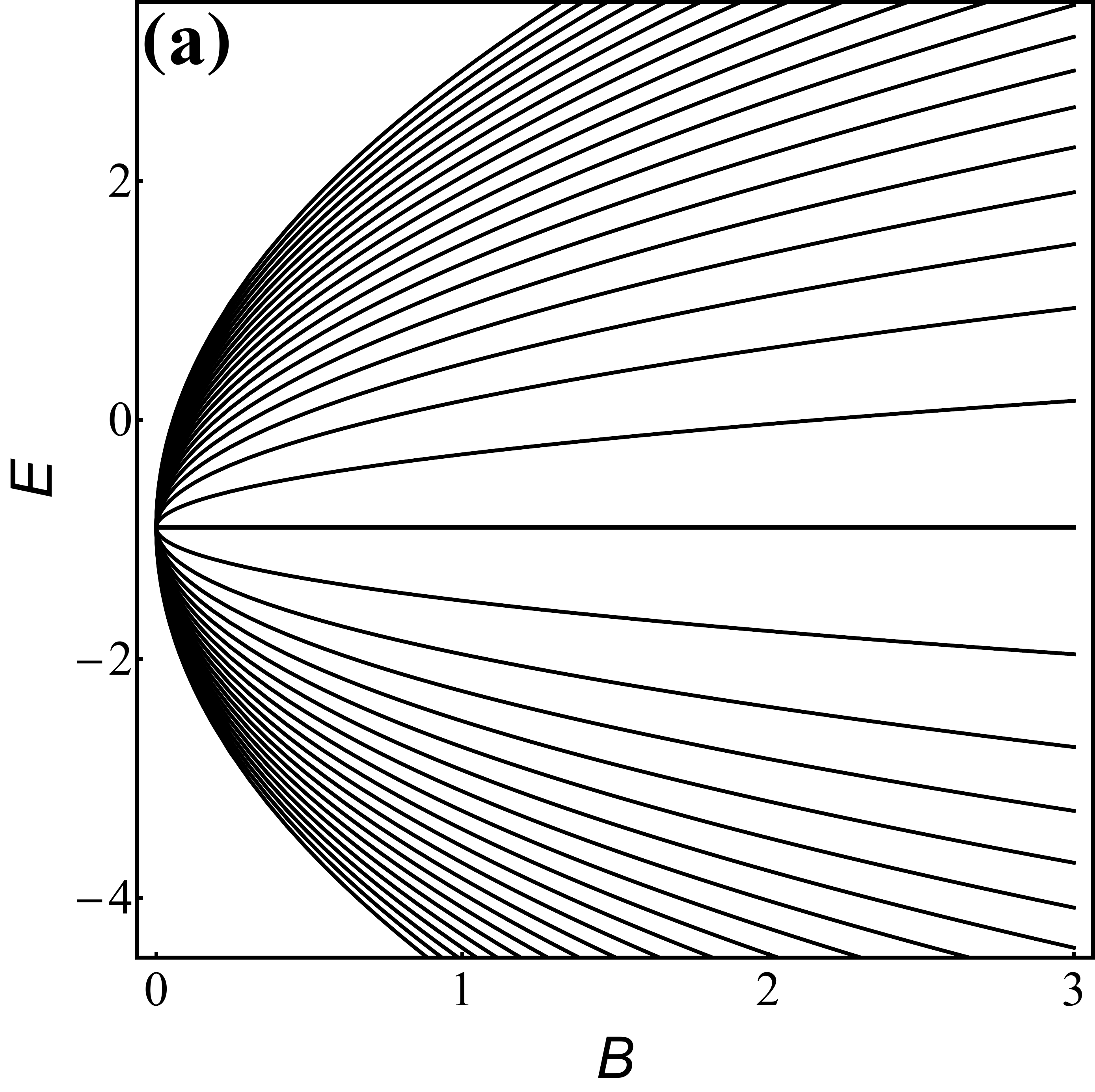}\label{LLLsa}}~~
  \subfigure{\includegraphics[width=1.7in]{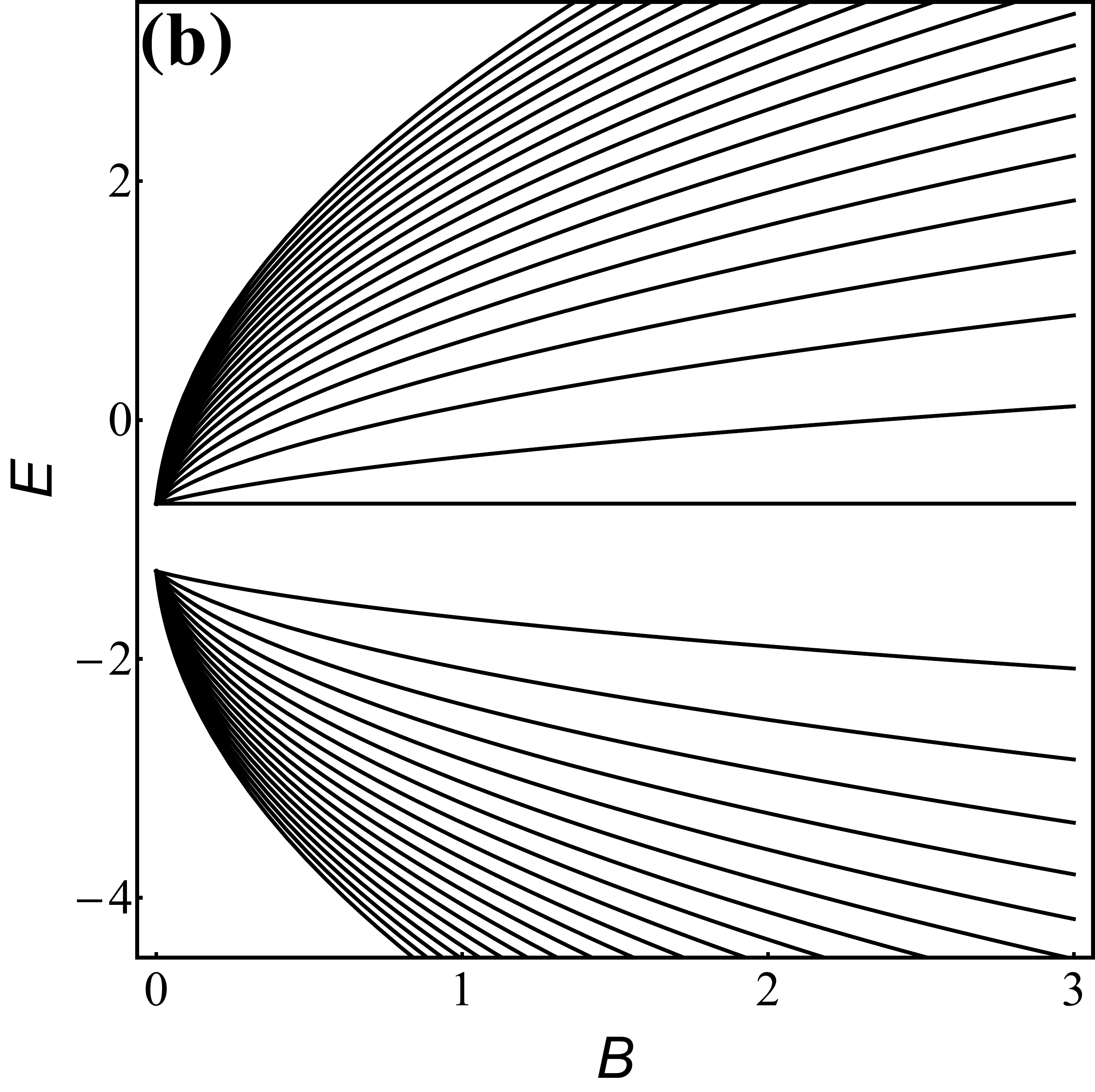}\label{LLLsb}}
  \caption{The Landau levels of TBCL as a function of magnetic B for (a) $\Delta=0$ and (b) $\Delta=-0.2$. In both figures, $k_z=0$ and $t_z=0.2$.}
  \label{FigLLLs}
\end{figure}
%%%%%%%%%%%%%%%%%%%%%%%%%%%%%%%%%%%%%%%%%%%
Retrieving the dropped term $\epsilon_{k_z}$, eventually, we have the LLs of Hamiltonian (\ref{Eq.MHS})
%%%%%%%%%%%%%%%%%%%%%%%%%%%%%%%%%%%%%%%%%%%
\begin{eqnarray}
E_{nk_z}=\epsilon_{k_z}, \quad \epsilon_{k_z} + \frac{v}{2l_B} \left(\delta + \lambda\sqrt{8n+4+{\delta}^2}\right).
\end{eqnarray}
When $\delta=0$ we obtain the LLs of the pseudospin-1 nodal line. The LLs are plotted in Fig.~\ref{FigLLLs}. Note that the flat level touches the upper levels or lower levels depending on the sign of $\Delta$.

Unlike in graphene where the zero-energy LL contributes to a half-integer anomaly in the Hall conductivity~\cite{V.P.Gusynin2005,M.O.Goerbig2011}, in the 2D pseudospin-1 system, the zero-energy flat band is nontopological and does not contribute to the Hall conductivity. Therefore, the Hall conductivity of the 2D pseudospin-$1$ system is given by $\sigma_H=2ne^2/h$ where the factor of $2$ is from spin~\cite{Z.Lan2011}. In the 3D pseudospin-1 nodal line system, when the Fermi level is in the 3D gap, the Hall conductivity is
%%%%%%%%%%%%%%%%%%%%%%%%%%%%%%%%%%%%%%%%%%%
\begin{eqnarray}
\sigma_{xy}=\frac{2ne^2}{h}\int\frac{dk_z}{2\pi}=\frac{2ne^2}{hc_0}
\end{eqnarray}
%%%%%%%%%%%%%%%%%%%%%%%%%%%%%%%%%%%%%%%%%%%
where $c_0$ is the lattice constant in the $z$-direction~\cite{Halperin1987,Jun-WonRhim2015,B.AndreiBernevig2007}. Experimentally, the three-dimensional quantum Hall effect has been observed in bulk zirconium pentatelluride (ZrTe$_5$) crystals~\cite{Fangdong.Tang2019}.

\subsection{Geometry dependent surface states}\label{subsec:surs}
\begin{figure}[t]
  \centering
  \subfigure{\includegraphics[width=1.5in]{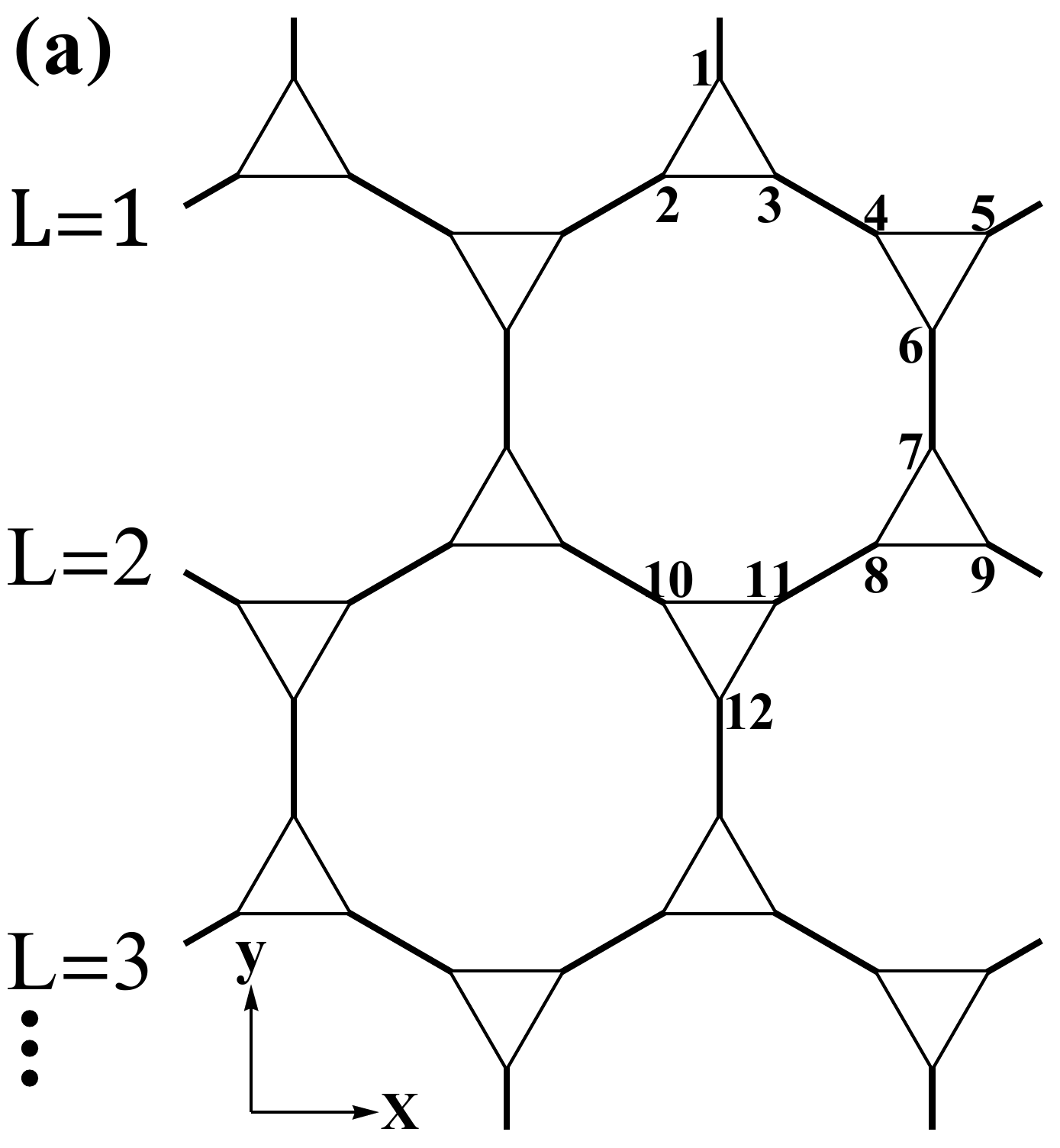}\label{Essa}}~~
  \subfigure{\includegraphics[width=1.7in]{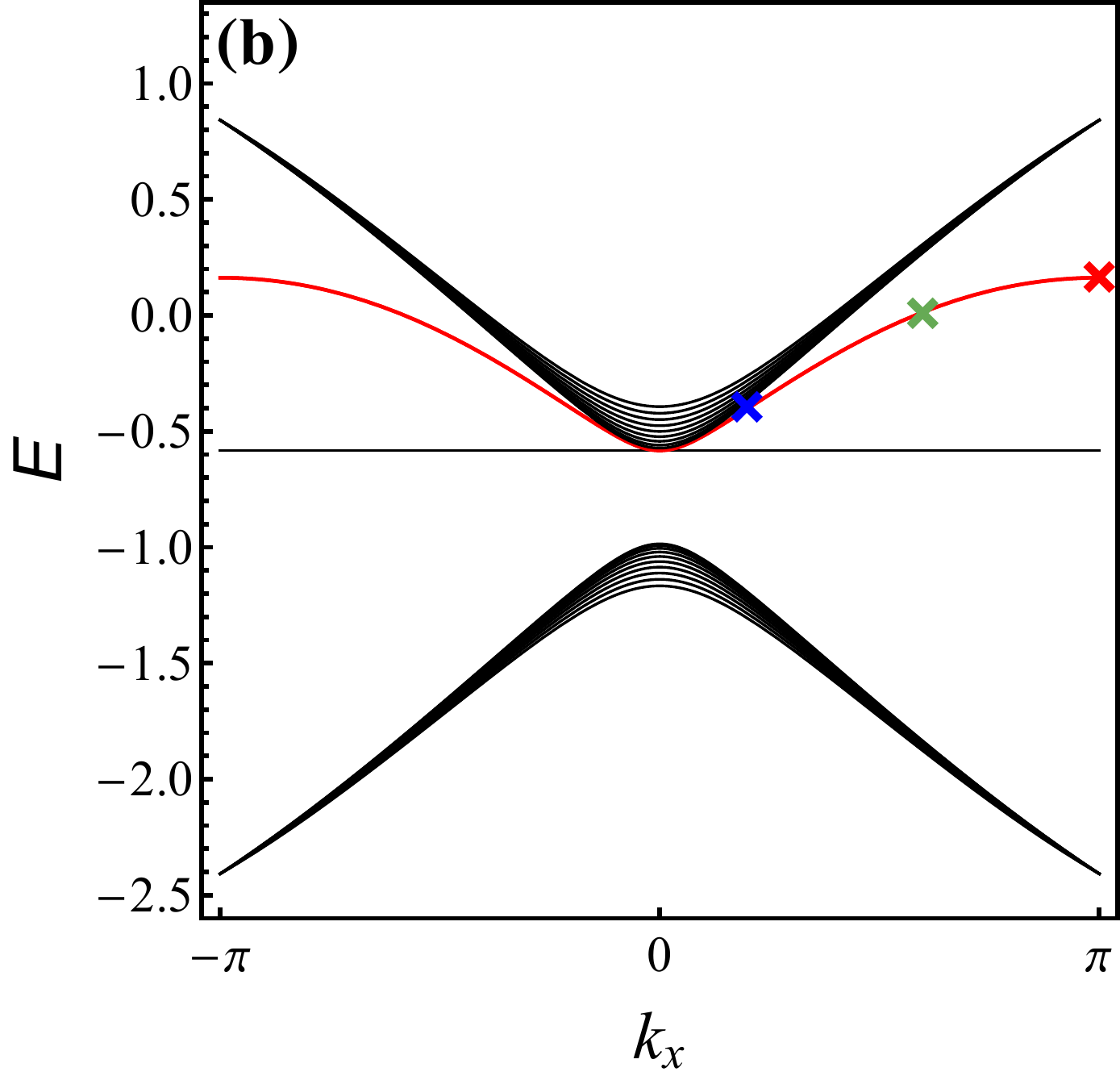}\label{Essb}}
  \subfigure{\includegraphics[width=1.6in]{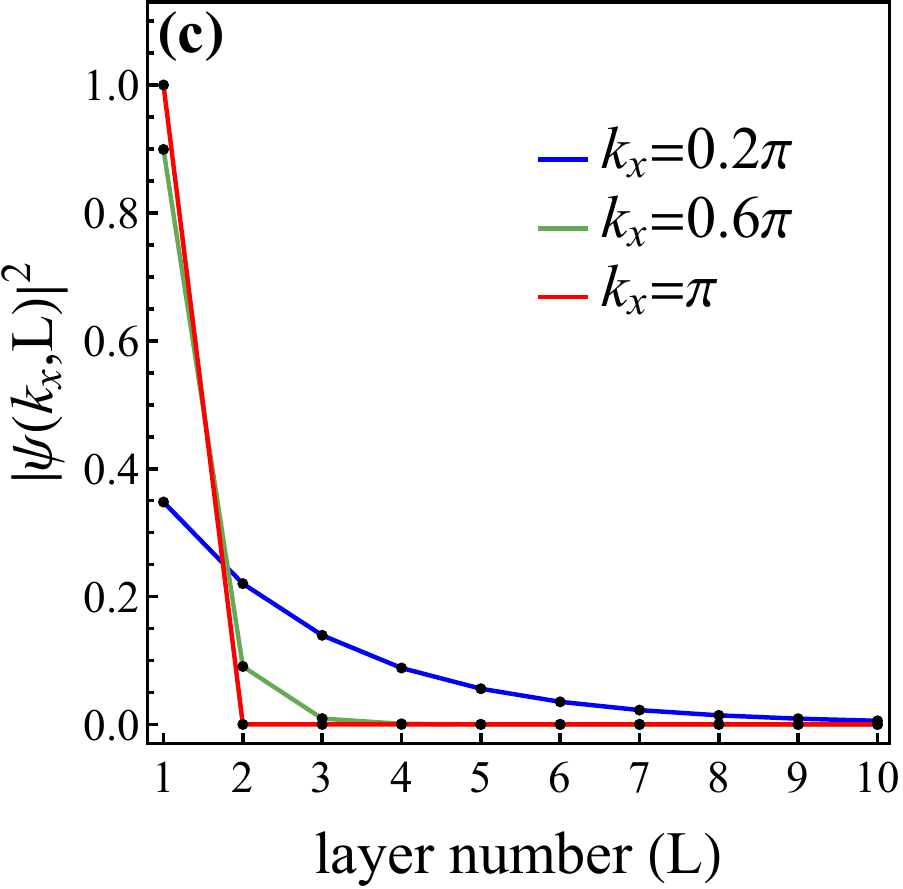}\label{Essc}}~~
  \subfigure{\includegraphics[width=1.6in]{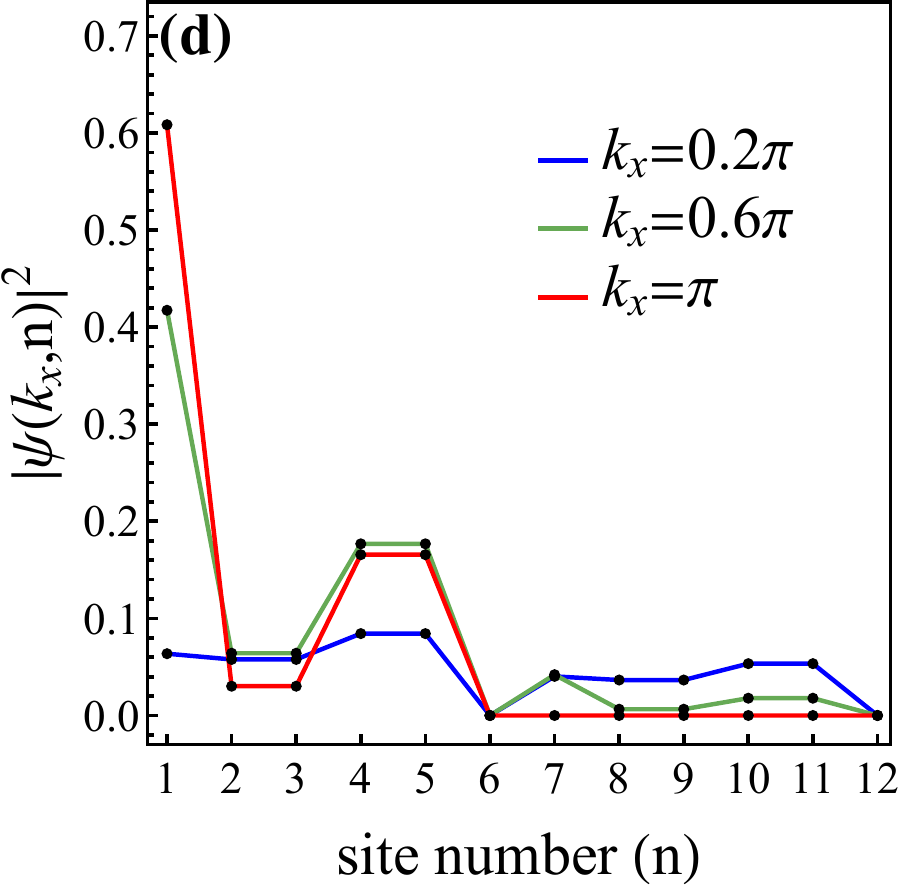}\label{Essd}}
  \caption{(a) Star lattice with layer number and site number labeled. (b) Band structure along $k_x$ for a slab finite in the $y$-direction, with $t^\prime=1.3t$, $t_z=0.2t$ and $k_z=\tfrac{\pi}{4}$. The surface bands are colored red. (c, d) The squared wave function of the surface state as a function of the layer number and the site number, respectively. Three states with different $k_x$'s are shown, corresponding to the three crosses in (b).} \label{FigESS}
\end{figure}

We choose a slab of the 3D star lattice to be infinite in the $x$- and $z$-directions but finite in the $y$-direction to study the surface states of the TBCL. The projection of the slab in the $xy$-plane is shown in Fig.~\ref{FigESS}(a), in which we label the layer number and the site number. For a fixed $k_z$, we calculate the band structure of the slab and have the results shown in Fig.~\ref{FigESS}(b). For clarity, we only show the bands near the TBCL. A couple of bands that are colored red deviate from the bulk bands which form the nodal line, and we recognize them as the bands of the surface states by inspecting the wave function of those states. In Figs.~\ref{FigESS}(c) and~\ref{FigESS}(d) we show the squared wave functions of the states corresponding to the crosses in Fig.~\ref{FigESS}(b) which are exponentially localized at the surface. $\psi(k_x,n)$ is the wave function, where $n$ is the site number shown in Fig.~\ref{FigESS}(c), and $|\psi(k_x,L)|^2$ is defined as the sum of $|\psi(k_x,n)|^2$ over $n$ in the $L$th layer. The surface states are non-topological and are dependent on the geometry of the lattice as well as the surface termination~\cite{Yang2014PRL}. For example, in the stacking of Lieb lattices where the TBCL also exist, there are no surface states. Significantly, upon achieving the pseudospin-1 nodal line by tuning parameters, the surface states are not affected qualitatively.

%%%%%%%%%%%%%%%%%%%%%%%%%%%%%%%%%%%%%%%%%%%
%%%%%%%%%%%%%%%%%%%%%%%%%%%%%%%%%%%%%%%%%%%

\subsection{FS nesting insensible to filling and spin density wave}\label{subsec:SDW}
We have presented the FS of the 3D star lattice at half filling in Fig.~\ref{FigSN}(a), which consists of two closed Fermi pockets and two flat pieces. Away from half filling, upon tuning the chemical potential $\mu$, one pocket becomes larger and the other becomes smaller, while the two flat pieces remain, whose positions depend on $\mu$. The two flat pieces are perfectly nested, with nesting vector ${\bf Q}= \left(0,0,2\arccos(\tfrac{\mu+0.5t}{2t_z})\right)$. As is well known, the FS nesting could induce density wave orders when interaction is taken into account~\cite{George.Gruner1994,Hiizu.Nakanishi1984}.

We consider a Hubbard model with Hamiltonian $\mathcal{H}=\mathcal{H}_0+\mathcal{H}_{z}+\mathcal{H}_I$, where $\mathcal{H}_0$ and $\mathcal{H}_z$ are given by Eq.~(\ref{Eq.LSH}) and Eq.~(\ref{Eq:Hz}), respectively, and $\mathcal{H}_I$ is the interacting Hamiltonian. So the Hubbard Hamiltonian can be written as
%%%%%%%%%%%%%%%%%%%%%%%%%%%%%%%%%%%%%%%%%%%
\begin{eqnarray}
\mathcal{H}=\sum_{i j \alpha \sigma}\left(t_{i j}c_{i \alpha \sigma}^\dagger c_{j \alpha \sigma}
+\mathrm{h.c.}\right) + U\sum_{i \alpha}n_{i \alpha \uparrow}n_{i \alpha \downarrow},
\end{eqnarray}
%%%%%%%%%%%%%%%%%%%%%%%%%%%%%%%%%%%%%%%%%%%
where $n_{i\alpha\sigma}=c_{i\alpha\sigma}^\dagger c_{i\alpha\sigma}$, $i$ and $j$ label two neighbour unit cells in 3D, $\alpha=1 , ... , 6$ label the six sites (orbitals) in each unit cell, and $U$ is on-site Coulomb interaction.

Taking the mean-field approximation, in momentum space, we obtain the following mean-field Hamiltonian
%%%%%%%%%%%%%%%%%%%%%%%%%%%%%%%%%%%%%%%%%%%
\begin{eqnarray}
\begin{split}
\mathcal{H}=\sum_{{\bf k} \sigma}E_{\bf k} c_{{\bf k} \sigma}^\dagger c_{{\bf k} \sigma}
&+ \tfrac{U}{2N} \sum_{{\bf k} {\bf k}^\prime \sigma}\Big[\langle c_{{\bf k} \sigma}^\dagger c_{{\bf k}+{\bf Q} \sigma}\rangle c_{{\bf k}^\prime \bar{\sigma}}^\dagger c_{{\bf k}^\prime-{\bf Q} \bar{\sigma}} \\
&+\langle c_{{\bf k}^\prime \bar{\sigma}}^\dagger c_{{\bf k}^\prime-{\bf Q} \bar{\sigma}}\rangle c_{{\bf k} \sigma}^\dagger c_{{\bf k}+{\bf Q} \sigma}\Big],
\end{split}
\label{Eq.MFH}
\end{eqnarray}
%%%%%%%%%%%%%%%%%%%%%%%%%%%%%%%%%%%%%%%%%%%
where $E_{\bf k}$ is the band energy and $N$ is the number of unit cells, $\bar{\sigma}$ is the opposite spin to $\sigma$, and we have dropped a constant term. Since both two Fermi pieces (see Fig.~\ref{FigSN}(a)) stem from the same energy band (corresponding to the lower flat band for each $k_z$), the nesting has no impact on other bands, hence, in Eq.~(\ref{Eq.MFH}) we consider only one band.

In the spirit of the nearly free electron approximation, in our work, we consider only ${\bf k}$ which nearby the two flat FS pieces (labelled by the subscripts $1$ and $2$ respectively). The nesting property of FS is given by $\epsilon_{1{\bf k}}=-\epsilon_{2{\bf k}}=\epsilon_{{\bf k}}$, where $\epsilon_{\bf k}$ is the energy measured from the Fermi energy and ${\bf k}$ is measured from the Fermi surface piece $1$ or $2$. We define real order parameters ${\Delta}_{\sigma}=\tfrac{U}{2N}\sum_{{\bf k}}\sigma\langle c_{1{\bf k} \sigma}^\dagger c_{2{\bf k} \sigma}\rangle=\sigma \Delta$. Then Eq.~(\ref{Eq.MFH}) becomes
%%%%%%%%%%%%%%%%%%%%%%%%%%%%%%%%%%%%%%%%%%%
\begin{equation}
\begin{split}
\mathcal{H}({\bf k})=&\sum_{{\bf k} \sigma}\epsilon_{\bf k}\left(c_{1{\bf k} \sigma}^\dagger c_{1{\bf k} \sigma} - c_{2{\bf k} \sigma}^\dagger c_{2{\bf k} \sigma}\right) \\
&+ \sum_{{\bf k} \sigma}{\Delta}_{\bar{\sigma}}\left(c_{1{\bf k} \sigma}^\dagger  c_{2{\bf k} \sigma} + c_{2{\bf k} \sigma}^\dagger  c_{1{\bf k} \sigma}\right),
\end{split}
\end{equation}
%%%%%%%%%%%%%%%%%%%%%%%%%%%%%%%%%%%%%%%%%%%
which can be diagonalized by the Bogoliubov transformation
%%%%%%%%%%%%%%%%%%%%%%%%%%%%%%%%%%%%%%%%%%%
\begin{equation}
\begin{split}
&\gamma_{1{\bf k} \sigma}=u_{\bf k}c_{1{\bf k} \sigma}-v_{\bf k}c_{2{\bf k} \sigma} \\
&\gamma_{2{\bf k} \sigma}=u_{\bf k}c_{2{\bf k} \sigma}+ v_{\bf k}c_{1{\bf k} \sigma},
\end{split}
\end{equation}
%%%%%%%%%%%%%%%%%%%%%%%%%%%%%%%%%%%%%%%%%%%
with the relation $u_{\bf k}^2+v_{\bf k}^2=1$. Making the off-diagonal terms obtained from the transformation vanishing, we get
%%%%%%%%%%%%%%%%%%%%%%%%%%%%%%%%%%%%%%%%%%%
\begin{eqnarray}
\mathcal{H}({\bf k})=\sum_{{\bf k} \sigma}\xi_{\bf k} \left(\gamma_{1{\bf k} \sigma}^\dagger \gamma_{1 {\bf k} \sigma} - \gamma_{2{\bf k} \sigma}^\dagger \gamma_{2{\bf k} \sigma}\right),
\end{eqnarray}
%%%%%%%%%%%%%%%%%%%%%%%%%%%%%%%%%%%%%%%%%%%
where $\gamma_{1{\bf k} \sigma}^\dagger$ ($\gamma_{2{\bf k} \sigma}^\dagger$) creates a quasi-particle near the FS piece $1$ ($2$) with spin $\sigma$. The quasiparticle energy spectrum $\xi_{\bf k}=\sqrt{\epsilon_{\bf k}^2+\Delta^2}$ demonstrates that a gap has developed near the Fermi level, from which the system has an energy gain and the ground state is a spin density wave.

\section{Pseudospin-1 vortex ring}\label{subsec:spins}

\subsection{The vortex ring model}\label{subsec:nr}
We study another system which possesses a triply degenerate nodal line, namely the pseudospin-1 vortex ring model. The pseudospin-$1/2$ vortex ring model has been investigated in Ref.~\onlinecite{Lih-KingLim2017}. In the following, we extend it to the pseudospin-1 case. The Hamiltonian reads
%%%%%%%%%%%%%%%%%%%%%%%%%%%%%%%%%%%%%%%%%%%
\begin{equation}
\begin{split}
H(k_x,k_y,k_z)=&-\tfrac{1}{m_z}k_xk_zS_x -\tfrac{1}{m_z}k_yk_zS_y \\
&+\tfrac{1}{2m_r}\left(k_x^2+k_y^2-k_z^2-\Delta\right)S_z,
\end{split}
\label{Eq.VRH}
\end{equation}
%%%%%%%%%%%%%%%%%%%%%%%%%%%%%%%%%%%%%%%%%%%
where $m_z$ and $m_r$ are parameters related to the Fermi velocity and are set to $1$ hereinafter. We choose the three spin-$1$ matrices to be
%%%%%%%%%%%%%%%%%%%%%%%%%%%%%%%%%%%%%%%%%%%
\begin{equation}
\begin{split}
&S_x=\tfrac{1}{\sqrt{2}}
\begin{pmatrix}
0 & 1 & 0 \\
1 & 0 & 1 \\
0 & 1 & 0
\end{pmatrix},
S_y=\tfrac{1}{\sqrt{2}}
\begin{pmatrix}
0 & -i & 0 \\
i & 0 & -i \\
0 & i & 0
\end{pmatrix}, \\
&S_z=
\begin{pmatrix}
1 & 0 & 0 \\
0 & 0 & 0 \\
0 & 0 & -1
\end{pmatrix},
\end{split}
\end{equation}
%%%%%%%%%%%%%%%%%%%%%%%%%%%%%%%%%%%%%%%%%%%
which are different from those in Sec.\ref{sec:starL}, since the diagonal matrix $S_z$ is more convenient for the subsequent calculations. This Hamiltonian describes a pseudospin-$1$ nodal ring structure in the $k_z=0$ plane with radius $k_0=\sqrt{\Delta}$ when $\Delta>0$. Tuning $\Delta$ towards $0$, the nodal ring shrinks into a point at $\Delta=0$, and for $\Delta<0$, two triply degenerate points emerge at ${\bf K}_{W\pm}=(0,0,\pm \sqrt{|\Delta|})$, that is the Dirac-Weyl (DW) phase. The vortex ring model respects an artificial symmetry: a mirror reflection with respect to the $z=0$ plane combined with an opposite parity of the orbitals under such a transformation (see Fig.~\ref{FigRL}). The Bloch Hamiltonian $H({\bf k})=\boldsymbol{h}({\bf k})\cdot \boldsymbol{S}$ thus transforms as $\boldsymbol{h}({\bf k})\to \left( -h_x(k_x,k_y,-k_z),-h_y(k_x,k_y,-k_z),h_z(k_x,k_y,-k_z) \right)$~\cite{Lih-KingLim2017}.
%%%%%%%%%%%%%%%%%%%%%%%%%%%%%%%%%%%%%%%%%%%
\begin{figure}[t]
  \centering
  \subfigure{\includegraphics[width=1.6in]{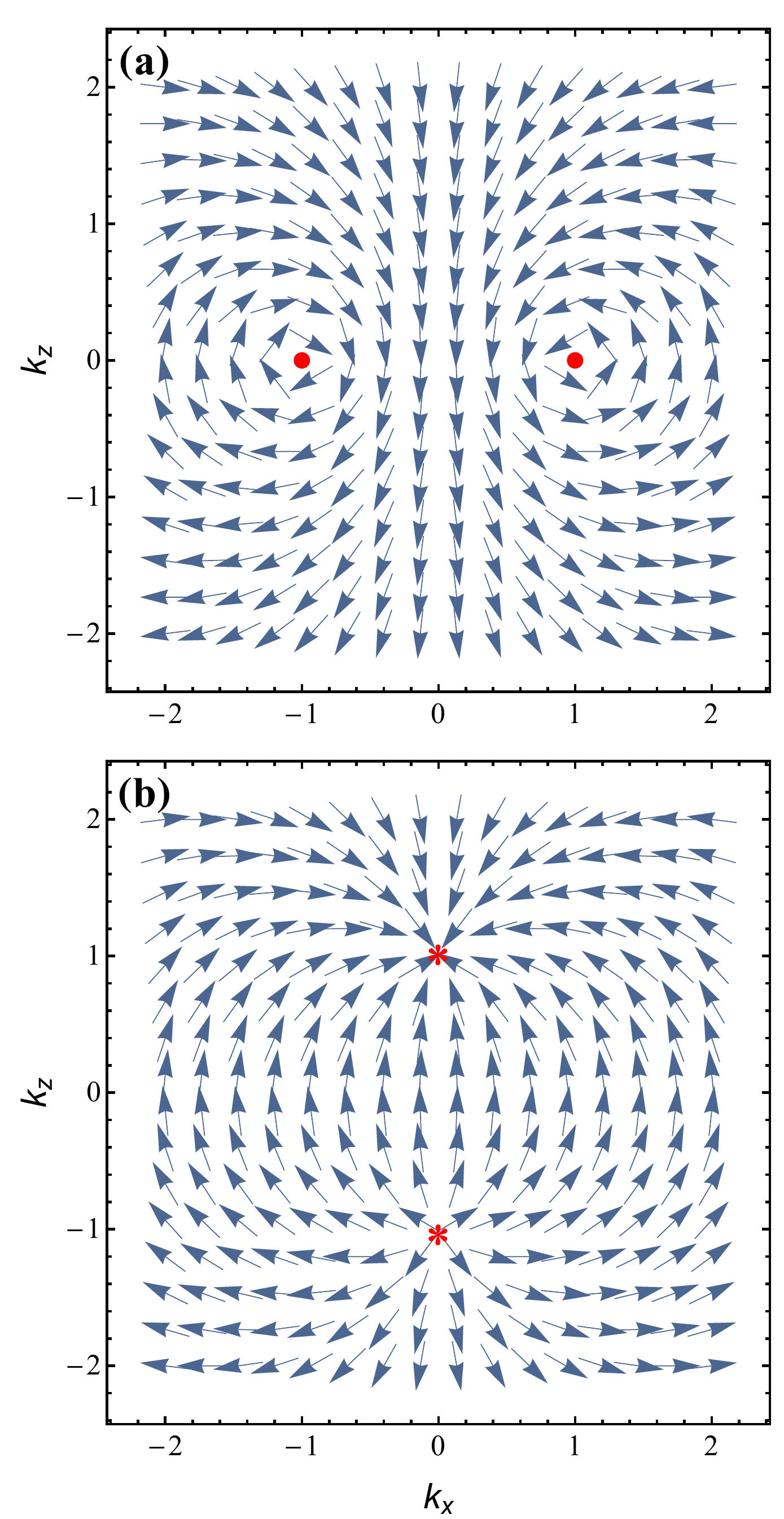}\label{cm}}~~~~
  \subfigure{\includegraphics[width=1.6in]{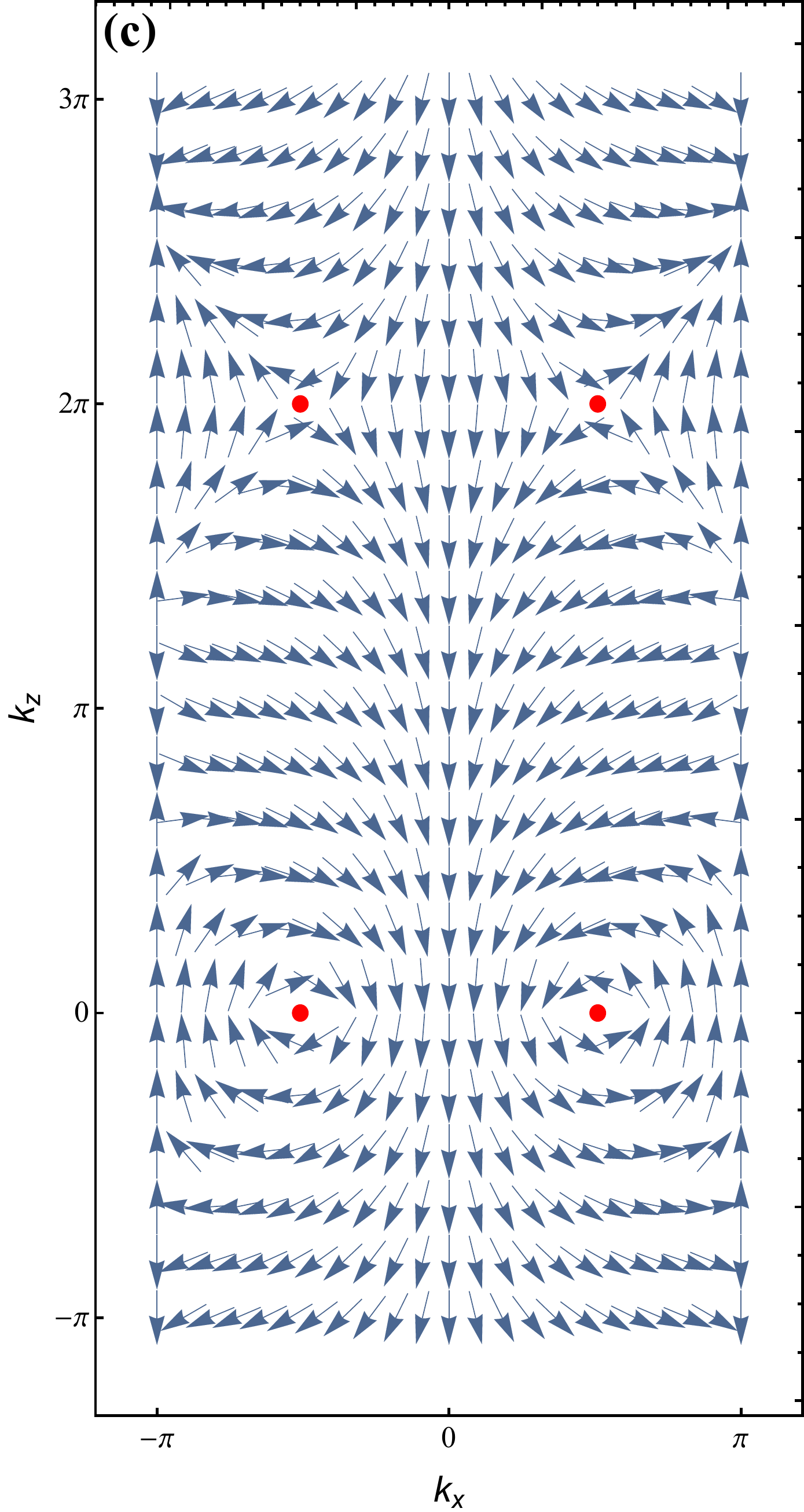}\label{tb}}
  \caption{Pseudospin textures in the $k_y=0$ plane for (a) the vortex ring phase in the continuous model, (b) the DW phase in the continuous model, and (c) the vortex ring phase in the tight-binding model. Red dots and asterisks represent the vortex ring and the triply degenerate points, respectively. The arrows implicate the projection of pseudospin-1 in $k_y=0$ plane, which is defined as $(\langle\phi|S_x|\phi\rangle, \langle\phi|S_z|\phi\rangle)$ where $|\phi\rangle$ is the eigenvector of the upper band.}\label{FigPSP}
\end{figure}
%%%%%%%%%%%%%%%%%%%%%%%%%%%%%%%%%%%%%%%%%%%

In Fig.~\ref{FigPSP}, we plot the pseudospin textures of the vortex ring and DW phases. Considering the rotational symmetry about the $k_z$-axis, we only present the results in the $k_y=0$ plane. For the vortex ring phase, two vortices with winding number 1 (with respect to the upper band) appear at $(\pm\sqrt{\Delta}, 0, 0)$, respectively, as shown in Fig.~\ref{FigPSP}(a). We also observe the Skyrmion structure in any $k_z\neq0$ plane. In Fig.~\ref{FigPSP}(b), as we can see, in the DW phase two triply degenerate points respectively serve as the source and drain, and the pseudospin flux flows from ${\bf K}_{W-}$ to ${\bf K}_{W+}$.
%%%%%%%%%%%%%%%%%%%%%%%%%%%%%%%%%%%%%%%%%%%
\begin{figure}[t]
  \centering\includegraphics[width=3.2in]{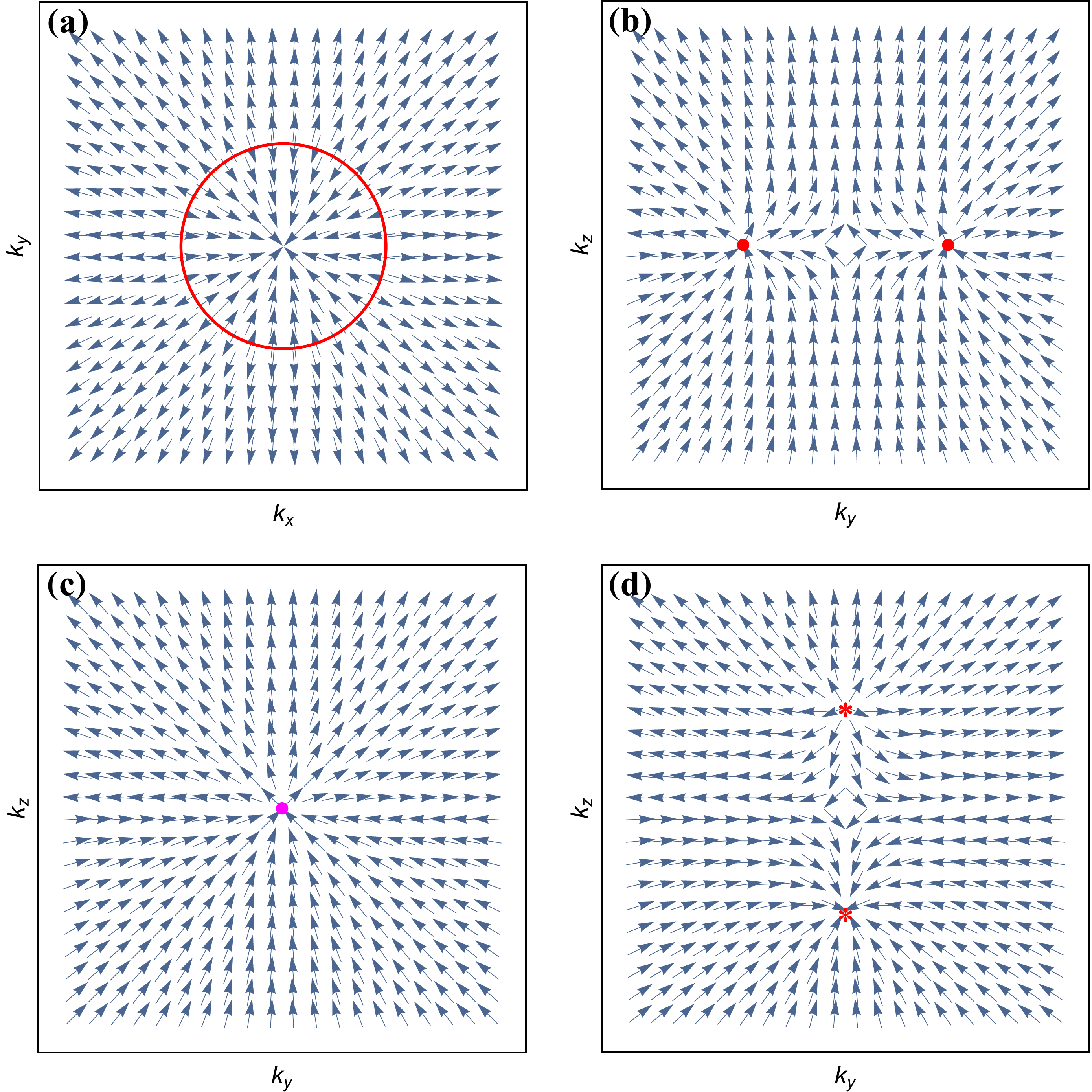}
  \caption{The normalized Berry curvature in (a) the vortex ring phase, in $k_z=0$ plane; (b) the vortex ring phase, in $k_x=0$ plane; (c) the critical point ($\Delta=0$), in $k_x=0$ plane; and (d) the DW phase, in $k_x=0$ plane. The red circle and dots denote the vortex ring, the red asterisks denote the triply degenerate points, and the magenta dot denotes the quadratic band touching.}\label{FigBC}
\end{figure}
%%%%%%%%%%%%%%%%%%%%%%%%%%%%%%%%%%%%%%%%%%%

We also depict the normalized Berry curvature of the highest band in Fig.~\ref{FigBC} for the vortex ring phase, the DW phase, and the critical point where $\Delta=0$. In the DW phase, the two triply degenerate points can be regarded as the source and drain of the Berry curvature, as one can see in Fig.~\ref{FigBC}(d), which is similar to Weyl semimetal phase in the pseudospin-1/2 case. The behavior of vortex ring in Berry curvature field can be understood by Fig.~\ref{FigBC}(c) where two triply degenerate points overlap ($\Delta=0$) and the three bands touch quadratically. We also calculate the planar Chern composition (PCC)~\cite{Lih-KingLim2017} which is the integration of the Berry curvature in a 2D cut of the first BZ. For the vortex ring phase, we obtain $C=2$ in different $k_z\neq0$ planes. In the DW case, in different $k_z$ planes we obtain $C=0$ when $|k_z|<\sqrt{|\Delta|}$ and $C=2$ for $|k_z|>\sqrt{|\Delta|}$. For both phases, we get $C=0$ in an arbitrary plane perpendicular to the $k_x$($k_y$)-axis that does not contain any node.

\subsection{Tight-binding realization and surface states}\label{subsec:TB}
Now we provide a tight-binding model corresponding to the continuous model and calculate the surface states. Assume the lattice constants are $a$ and $b$ in the $z$ and $x$ ($y$) directions, respectively. The tight-binding Hamiltonian is $H_{TB}({\bf k})=\boldsymbol{h}({\bf k})\cdot \boldsymbol{S}$ where
%%%%%%%%%%%%%%%%%%%%%%%%%%%%%%%%%%%%%%%%%%%
\begin{equation}
\begin{split}
h_x({\bf k})=&-\tfrac{2\hbar^2}{ab}\tfrac{\alpha \pi}{\sin{(\alpha\pi)}}\sin{(bk_x)}\sin{(ak_z/2)},   \\
h_y({\bf k})=&-\tfrac{2\hbar^2}{ab}\tfrac{\alpha \pi}{\sin{(\alpha\pi)}}\sin{(bk_y)}\sin{(ak_z/2)},   \\
h_z({\bf k})=&\tfrac{\hbar^2}{b^2}\tfrac{\alpha \pi}{\sin{(\alpha\pi)}} \Big(2-\cos{(bk_x)}-\cos{(bk_y)}   \\
& +\cos{(\alpha\pi)}-r\Big) -\tfrac{\hbar^2}{a^2} \left(1-\cos{(ak_z)} \right).
\end{split}
\label{Eq.TBH}
\end{equation}
%%%%%%%%%%%%%%%%%%%%%%%%%%%%%%%%%%%%%%%%%%%
In the tight-binding model, the vortex ring is not circular anymore due to the underlying lattice. At small $\alpha$, expanding $H_{TB}$ near the nodal ring ($k_z=0$ and $k_r=\alpha \pi /b$), we can recover the Hamiltonian (\ref{Eq.VRH}) when $r=1$. Noteworthily, no real $\alpha$ can turn the Hamiltonian (\ref{Eq.TBH}) into the DW phase when $r=1$. Therefore, to discuss the two phases both in real $\alpha$ space, we introduce a tunable parameter $r$.

We characterize the pseudospin texture of the tight-binding Hamiltonian in Fig.~\ref{FigPSP}(c). When $k_z$ locates at the interval ($-\pi$, $\pi$) we recover the texture in the continuous model (Fig.~\ref{FigPSP}(a)). However, it has period $4\pi$ in the $k_z$-direction because of the existence of the $\sin{(ak_z/2)}$ term, so the picture in the interval $k_z\in(\pi,3\pi)$ has no counterpart in the continuous model.
%%%%%%%%%%%%%%%%%%%%%%%%%%%%%%%%%%%%%%%%%%%
\begin{figure}[t]
  \centering\includegraphics[width=3.4in]{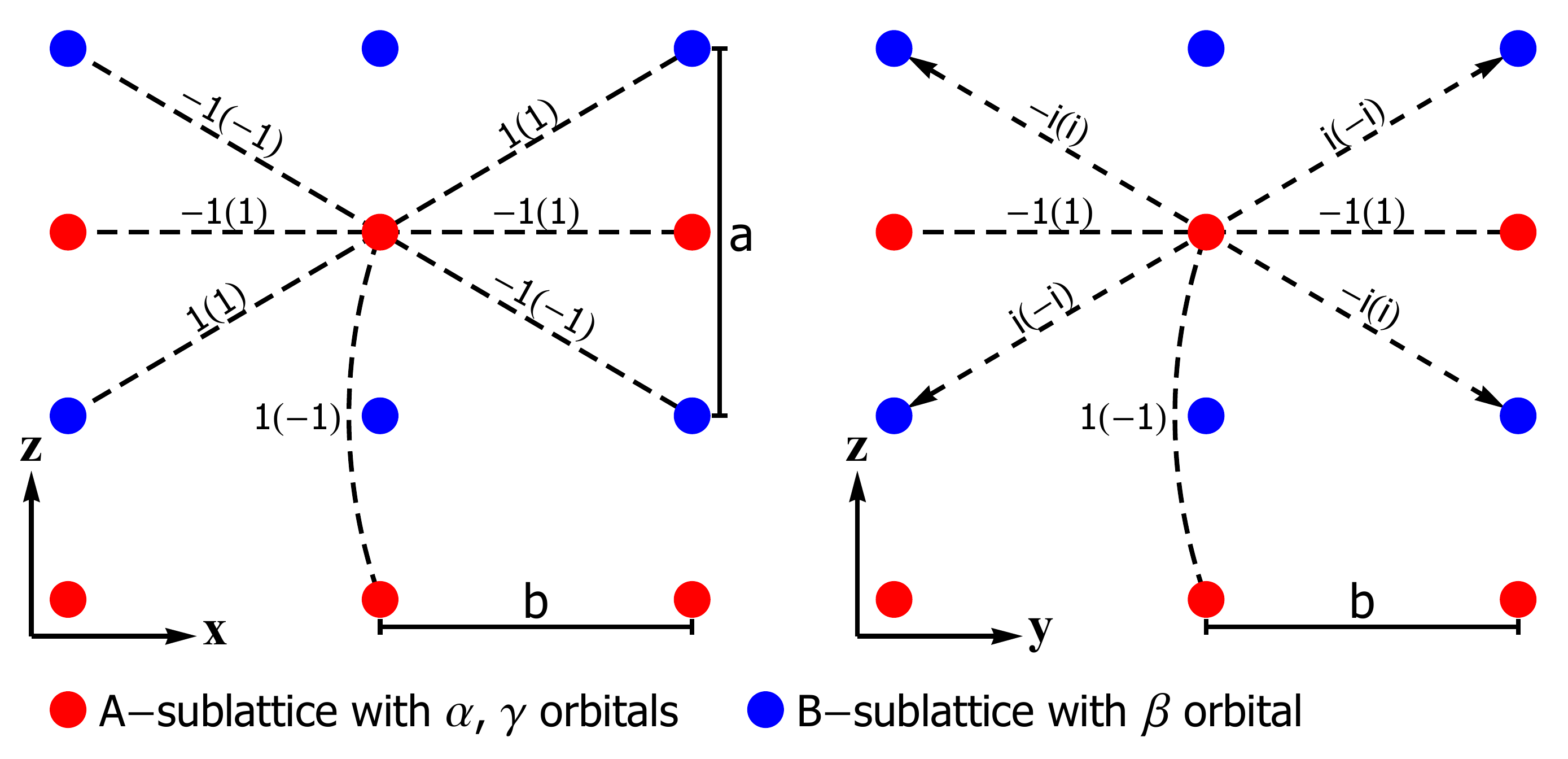}
  \caption{Schematic of the hoppings in the tight-binding Hamiltonian in the $xz$- (left) and $yz$-plane (right). The lattice consists of two sublattices A and B, and there exist three orbitals in each unit cell. The parameters outside (in) the brackets on dashed lines denote the normalized amplitudes of the hopping processes which $\alpha$ ($\gamma$) orbital participates in. $a$ and $b$ are the lattice constants along $z$ and $x$ ($y$) directions, respectively.}\label{FigRL}
\end{figure}
%%%%%%%%%%%%%%%%%%%%%%%%%%%%%%%%%%%%%%%%%%%

In real space, the tight-binding lattice contains two sublattices A and B, as shown in Fig.~\ref{FigRL}. Each site of A-sublattice contains two orbitals ($\alpha$ and $\gamma$) and each site of B-sublattice contains one ($\beta$). There is no hopping between neighbour $\beta$ orbitals in our model. Note that there exist complex hopping amplitudes in the $yz$-plane, so that the amplitudes of the reversed hopping possesses are the complex conjugate.

\begin{figure}[t]
  \centering
  \subfigure{\includegraphics[width=1.6in]{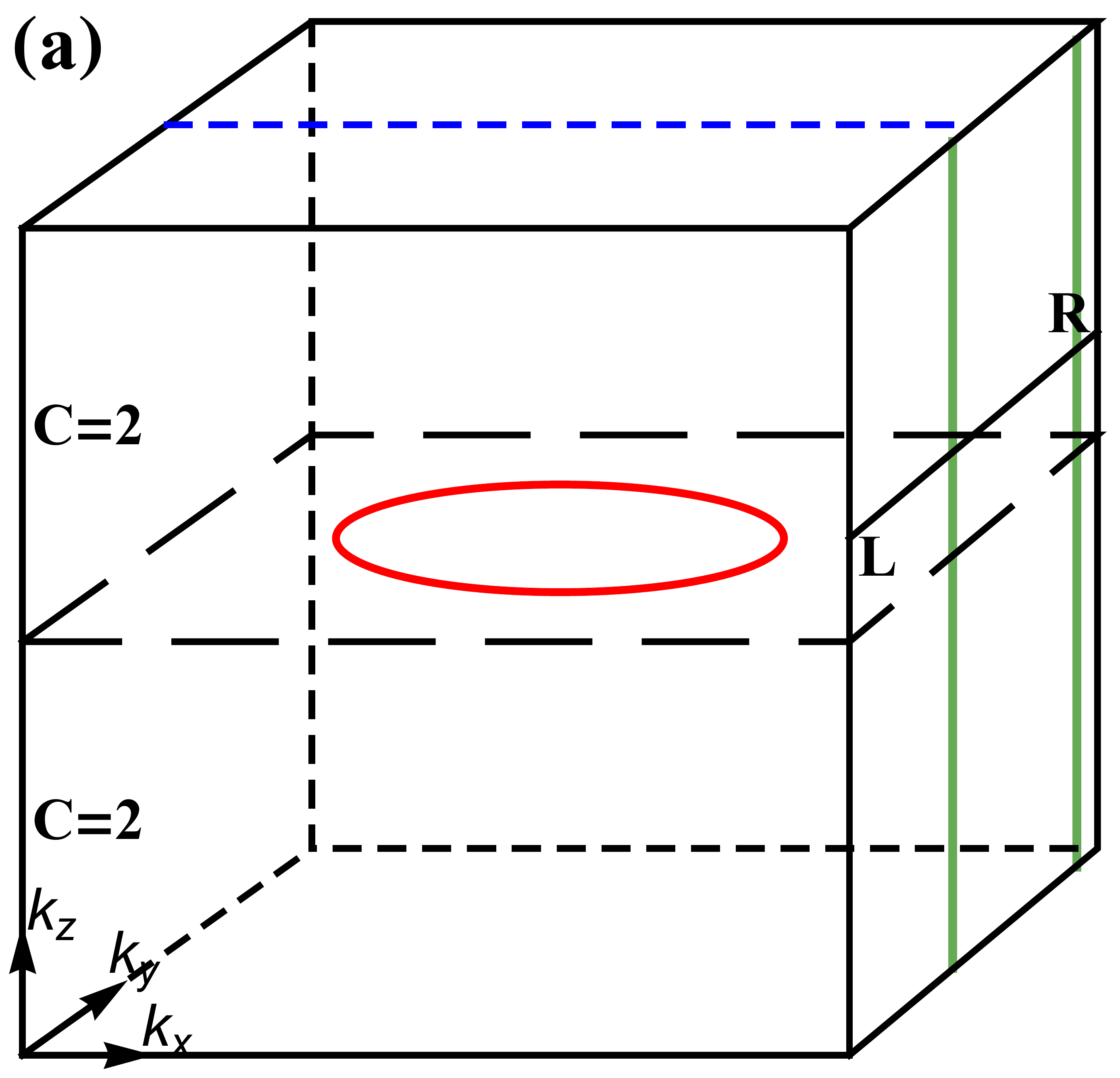}\label{Fasa}}~~
  \subfigure{\includegraphics[width=1.6in]{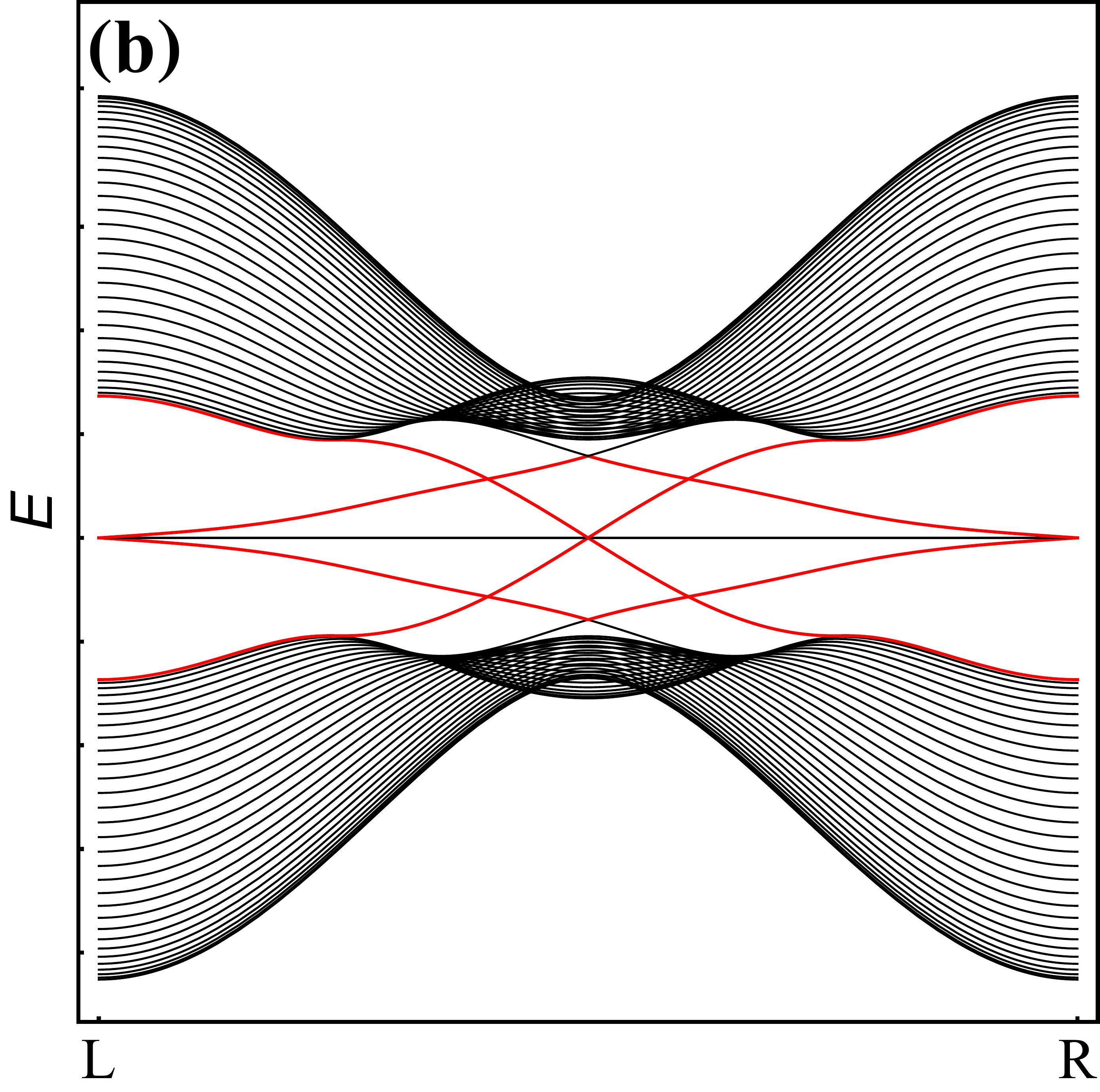}\label{Fasb}}
  \subfigure{\includegraphics[width=1.6in]{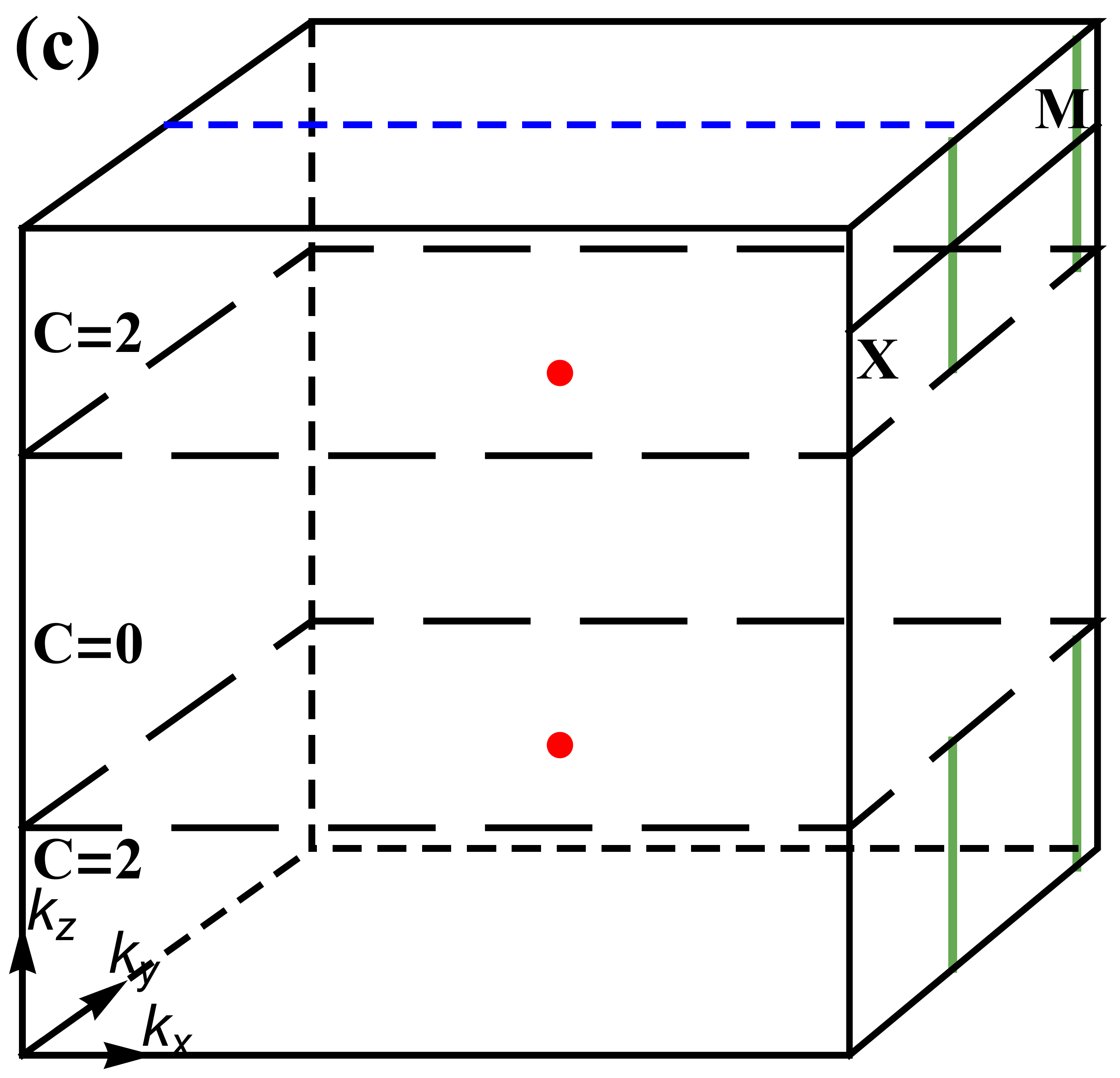}\label{Fasc}}~~
  \subfigure{\includegraphics[width=1.6in]{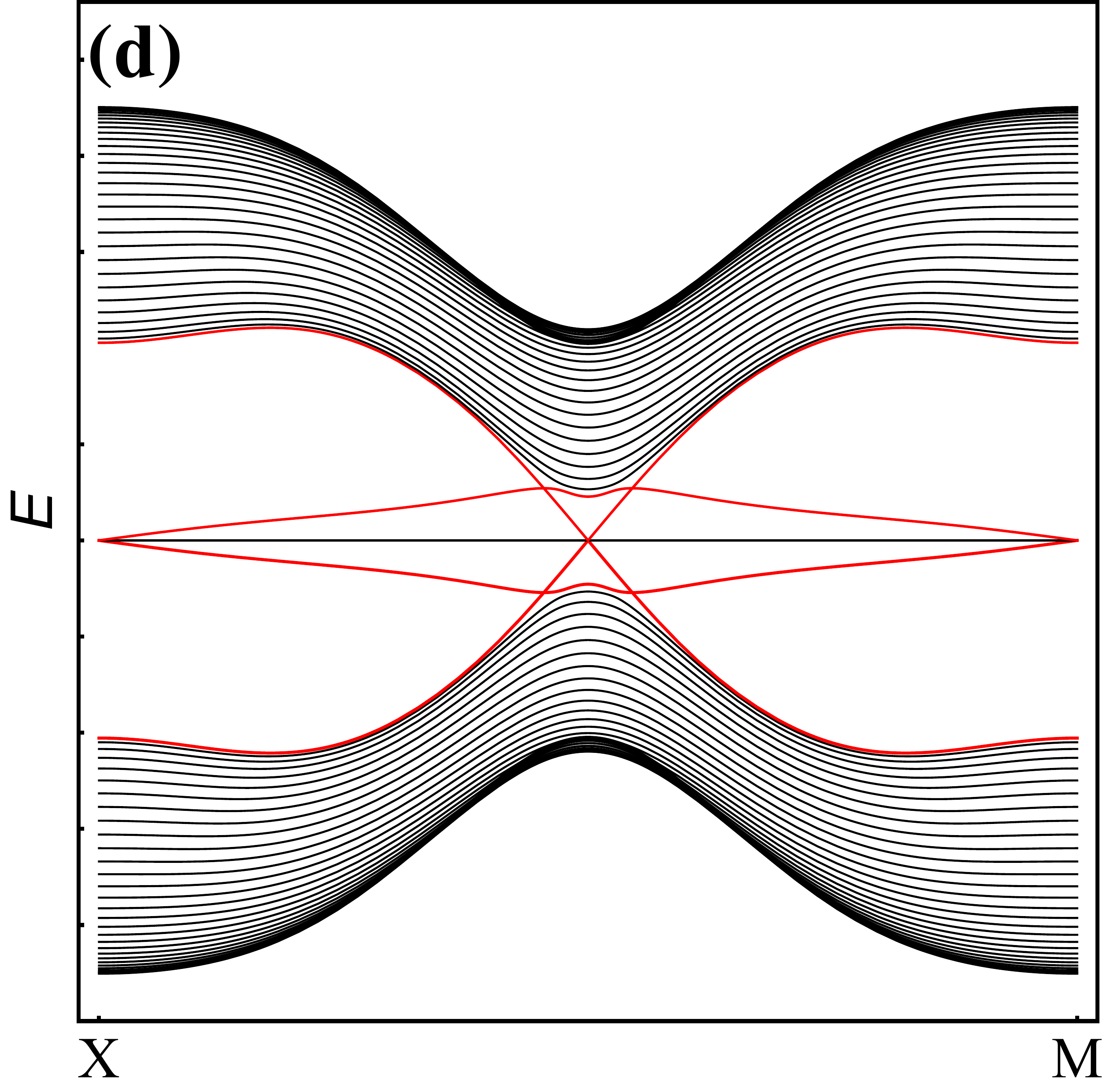}\label{Fasd}}
  \caption{Surface band structures of the tight-binding model in slab geometries (right) and the associated chiral Fermi arcs (left). The slab is finite in the $x$-direction. (a) and (b) are for the vortex ring phase, and (c) and (d) are for the DW phase. The red lines in (b) and (d) represent the chiral surface states, and the green lines in (a) and (c) indicate the chiral Fermi arcs on the surface BZ.}\label{FigSSR}
\end{figure}

We calculate the band structures of the tight-binding model in a slab which is infinite in the $y$- and $z$-directions but finite in the $x$-direction. In the vortex ring phase with $\alpha=0.5$ and $r=1$, for an arbitrary $k_z\neq0$ plane, two chiral surface states on each surface cross the bulk energy gaps. On the surface Brillouin zone parallel to $k_z$, these chiral surface states form two Fermi arcs which wrap around the full surface BZ, as shown in Figs.~\ref{FigSSR}(a) and~\ref{FigSSR}(b). This gives rise to a $3D$ quantum anomalous Hall effect with a maximal Hall conductivity $\sigma_{xy}^{3D}=(2e^2/2\pi h)(2\pi/a)$, where $(2\pi/a)$ is the magnitude of the primitive reciprocal vector along $k_z$. For the DW phase where we set $\alpha=0.25$ and $r=0.5$, when $|k_z|>\sqrt{|\Delta|}$, two chiral surface bands on each surface cross the gaps. However, there are no topological surface states in the interval $|k_z|<\sqrt{|\Delta|}$, as shown in Figs.~\ref{FigSSR}(c) and~\ref{FigSSR}(d). Consequently, the $3D$ Hall conductivity in DW phase is $\sigma_{xy}^{3D}=(2e^2/2\pi h)(2\pi/a-\sqrt{4|\Delta|})$. We also consider a slab finite in the $z$-direction, whose lower and upper surfaces are both A-sublattices, and find no chiral surface states in such a slab both for the vortex ring phase and the DW phase (along the blue dashed lines in Figs.~\ref{FigSSR}(a) and~\ref{FigSSR}(c)).
%%%%%%%%%%%%%%%%%%%%%%%%%%%%%%%%%%%%%%%%%%%
%%%%%%%%%%%%%%%%%%%%%%%%%%%%%%%%%%%%%%%%%%%

\subsection{Landau level structure}\label{subsec:LLs}
To obtain the LL structure of pseudospin-$1$ vortex ring system in a magnetic field ${\bf B}$ in the $z$-direction, we continue to use the above mentioned vector potential ${\bf A}$ as well as the ladder operators. Then we get the Hamiltonian in a magnetic field
%%%%%%%%%%%%%%%%%%%%%%%%%%%%%%%%%%%%%%%%%%%
\begin{eqnarray}
H_{B}=
\begin{pmatrix}
E_B^2(\hat{a}^\dagger \hat{a}+\tfrac{1}{2})-\tfrac{\delta}{2} & -P_zE_B\hat{a} & 0 \\
-P_zE_B\hat{a}^\dagger & 0 & -P_zE_B\hat{a} \\
0 &-P_zE_B\hat{a}^\dagger & \tfrac{\delta}{2}-E_B^2(\hat{a}^\dagger \hat{a}+\tfrac{1}{2})
\end{pmatrix},
\end{eqnarray}
%%%%%%%%%%%%%%%%%%%%%%%%%%%%%%%%%%%%%%%%%%%
where $E_B=\sqrt{eB\hbar}$, $P_z=\hbar k_z$ and $\delta=P_z^2+\Delta$.

Due to the existence of the $S_z$ term, $H_B$ cannot be diagonalized analytically for an arbitrary $n$. In order to obtain the LLs, we use a numerical method~\cite{YongXu2017}. Let the eigenvalues and eigenstates of $H_B$ be $E_n$ and $\psi_n=(\alpha_n, \beta_n, \gamma_n)^T$, respectively, where $\beta_n=b_n|n\rangle$, i.e. it satisfies $\hat{a}^\dagger \hat{a}\beta_n=n\beta_n$. Making use of the eigenequation $H_B\psi_n=E_n\psi_n$ and the relations $\hat{a}^\dagger|n\rangle=\sqrt{n+1}|n+1\rangle$ and $\hat{a}|n\rangle=\sqrt{n}|n-1\rangle$, the form of the eigenstates can be restricted as
%%%%%%%%%%%%%%%%%%%%%%%%%%%%%%%%%%%%%%%%%%%
\begin{eqnarray}
\psi_n=
\begin{pmatrix}
a_n|n-1\rangle \\
b_n|n\rangle \\
c_n|n+1\rangle
\end{pmatrix},
\end{eqnarray}
%%%%%%%%%%%%%%%%%%%%%%%%%%%%%%%%%%%%%%%%%%%
where $a_n$, $b_n$ and $c_n$ are the normalization factors. We first consider the $n=-1$ case, and obtain $\psi_{-1}=(0, 0, |0\rangle)^T$ and $E_{-1}=(\delta-E_B^2)/2$. When $n=0$, we get $E_{0,\pm}=\tfrac{1}{4}\left(-3E_B^2+\delta \pm \sqrt{9E_B^4+2E_B^2(8P_z^2-3\delta)+\delta^2} \right)$ and $\psi_{0,\pm}=(0, b_1|0\rangle, c_1|1\rangle)^T$ where $b_1=-P_zE_B/\sqrt{P_z^2E_B^2+E_{0,\pm}^2}$ and $c_1=E_{0,\pm}/\sqrt{P_z^2E_B^2+E_{0,\pm}^2}$. For $n>0$, the LLs are given by diagonalizing the Hamiltonian
%%%%%%%%%%%%%%%%%%%%%%%%%%%%%%%%%%%%%%%%%%%
\begin{eqnarray}
H_n=
\begin{pmatrix}
E_B^2(n-\tfrac{1}{2})-\tfrac{\delta}{2} & -E_BP_z\sqrt{n} & 0 \\
-E_BP_z\sqrt{n} & 0 & -E_BP_z\sqrt{n+1} \\
0 & -E_BP_z\sqrt{n+1} & \tfrac{\delta}{2}-E_B^2(n+\tfrac{3}{2})
\end{pmatrix}.
\label{Eq.LLH}
\end{eqnarray}
%%%%%%%%%%%%%%%%%%%%%%%%%%%%%%%%%%%%%%%%%%%

%%%%%%%%%%%%%%%%%%%%%%%%%%%%%%%%%%%%%%%%%%%
\begin{figure}[t]
  \centering\includegraphics[width=3.4in]{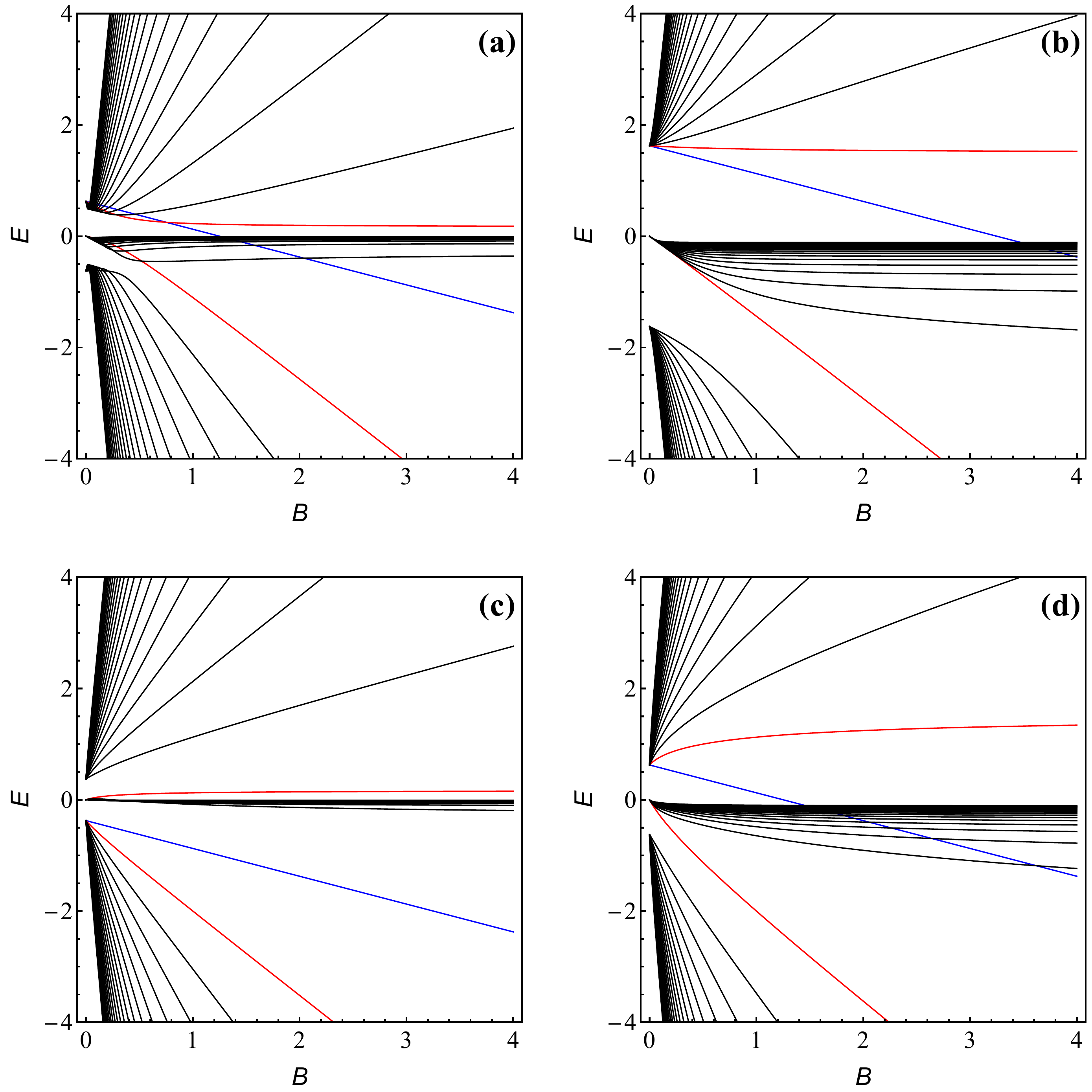}
  \caption{LLs of pseudospin-$1$ vortex ring model. (a) and (b) illustrate the vortex ring case with $\Delta=1$ at $k_z=0.5$ and $k_z=1.5$, respectively. (c) and (d) illustrate the DW case with $\Delta=-1$ at $k_z=0.5$ and $k_z=1.5$, respectively. Blue lines denote the LL $E_{-1}$  and red lines denote $E_{0,\pm}$.}\label{FigLL}
\end{figure}
%%%%%%%%%%%%%%%%%%%%%%%%%%%%%%%%%%%%%%%%%%%

We plot LLs of Hamiltonian (\ref{Eq.LLH}) as a function of the magnetic field in Fig.~\ref{FigLL}. Different from the pseudospin-$1$ nodal line system (the 3D star lattice), due to the existence of the $S_z$ term, the degeneracy of the flat band is lifted, leading to a series of LLs. A particular LL whose energy is linearly dependent on the magnetic field always exists, corresponding to $E_{-1}=(\delta-E_B^2)/2$. In the vortex ring phase, the separation between the middle group of LLs and the other groups increases with $k_z$, and the LLs $E_{-1}$ and $E_{0,+}$ always emit from the upper group while $E_{0,-}$ from the middle group, as shown in Fig.~\ref{FigLL}(a) and Fig.~\ref{FigLL}(b). In particular, the $n=-1$ state is a particlelike state at small fields, but transmutes into a holelike state at sufficiently large fields. In the DW phase, when $|k_z|<\sqrt{|\Delta|}$, $E_{0,+}$ emits from the middle group while $E_{-1}$ and $E_{0,-}$ from the lower group, as shown in Fig.~\ref{FigLL}(c); and when $|k_z|>\sqrt{|\Delta|}$, $E_{0,+}$ and $E_{-1}$ emit from the upper group while $E_{-1}$ from the middle group, as shown in Fig.~\ref{FigLL}(d).

\subsection{Fermion doubling of pseudospin-$1$ vortex lines}\label{subsec:dnl}
In the pseudospin-$1$ vortex ring Hamiltonian (\ref{Eq.VRH}), if we drop $k^2_y$ in the $S_z$-term, we obtain the Hamiltonian
%%%%%%%%%%%%%%%%%%%%%%%%%%%%%%%%%%%%%%%%%%%
\begin{eqnarray}
H_{L}({\bf k})=-k_xk_zS_x - k_yk_zS_y + \frac{1}{2}(k_x^2-k_z^2-\Delta)S_z.
\label{Eq.HL0}
\end{eqnarray}
%%%%%%%%%%%%%%%%%%%%%%%%%%%%%%%%%%%%%%%%%%%
Compared with the vortex ring model, this Hamiltonian presents two straight vortex lines along $k_x=\pm\sqrt{\Delta}$ in the $k_z=0$ plane. In an arbitrary $k_z\neq0$ plane, we obtain the PCC $C=2$. Expanding $H_{L}$ near the two vortex lines, we get
%%%%%%%%%%%%%%%%%%%%%%%%%%%%%%%%%%%%%%%%%%%
\begin{eqnarray}
H_{L,\pm}({\bf k})=-k_xk_zS_x - k_yk_zS_y \pm \delta\left(k_x\mp\delta\right)S_z.
\label{Eq.HL1}
\end{eqnarray}
%%%%%%%%%%%%%%%%%%%%%%%%%%%%%%%%%%%%%%%%%%%
We calculate the PCC of $H_{L,\pm}$ and get $C=1$ in any $k_z\neq0$ plane. Therefore, each vortex line contributes half Chern number of the PCC of $H_{L}$. That is reminiscent of the graphene system, in which each of the two Dirac points contributes Chern number $1/2$, making the whole system host Chern number $1$ when time reversal symmetry is broken, and such a Dirac point cannot singly exist due to the fermion-doubling theorem~\cite{H.B.Nielsen1981}. In a similar way, the vortex line described by Hamiltonian (\ref{Eq.HL1}) cannot singly exist in a lattice model if it is the only nodal feature of the system. The presence of two vortex lines can either make the PCC with $C=2$, or yield a topologically trivial system with $C=0$.

\section{summary}\label{subsec:sum}
We have investigated two types of triply degenerate nodal lines. For the first type, we use the 3D $AA$-stacked star lattice as an example, in which the band structures present a TBCL. The TBCL includes a symmetry-protected quadratic band-crossing line and a non-degenerate band and can form a triply degenerate (pseudospin-1) nodal line by fine tuning. We derived an effective Hamiltonian to describe the pseudospin-1 nodal line, which can also be used to describe the TBCL by considering a perturbation. In this way, We studied the splitting into Weyl nodal lines, the surface band structures, the 3D quantum Hall effect in a strong magnetic field and the instability to the SDW state due to FS nesting. Specifically, unlike in other systems where SDW only occurs at half filling, the SDW here can occur as long as the middle band of the pseudospin-1 band structure is partially filled due to the existence of two flat FS pieces in this range of filling.

The second type is a pseudospin-$1$ vortex ring model. In this model, by continuously tuning a parameter $\Delta$ from positive to negative, the system changes from the nodal ring phase to the DW phase. We also provided a tight-binding realization on a lattice, in which the chiral surface states form two Fermi arc on the surface BZ. In the nodal ring phase, the surface Fermi arcs wrap around the full surface BZ so that the model exhibits 3D quantum anomalous Hall effect with a maximal Hall conductivity. We also obtained the LLs and characterized the pseudospin structures and the Berry curvature in both the nodal ring and the DW phases. If we stretch the ring such that it is not closed in the first BZ, we can get two open nodal lines. In the continuous model, these two nodal lines give the PCC $C=2$ at an arbitrary $k_z$ plane except for $k_z=0$. We also found that each one of these two nodal lines can exist singly in the continuous model and contributes $C=1$ at different $k_z\neq 0$ planes, but it cannot exist on its own in the lattice model if it is the only nodal feature of the system.

\acknowledgments
This work was supported by NKRDPC-2017YFA0206203, NKRDPC-2018YFA0306001, NSFC-11974432, NSFG-2019A1515011337, National Supercomputer Center in Guangzhou, and Leading Talent Program of Guangdong Special Projects.

\bibliography{nodalline}

\bibliographystyle{apsrev4-1}

\end{document}